\documentclass[pre,amsmath,amssymb,showpacs]{revtex4-1}

\usepackage{graphicx}
\usepackage{bm}
\usepackage{amssymb}
\usepackage{colordvi}
\usepackage{color}     
\usepackage{amsbsy}
\usepackage{amsmath}
\usepackage{amsbsy}
\usepackage{latexsym}

\newcommand{\ba}{\mathbf{a}}
\newcommand{\bb}{\mathbf{b}}

\newcommand{\bM}{\mathbf{M}}

\newcommand{\bh}{\mathbf{h}}
\newcommand{\bH}{\mathbf{H}}
\newcommand{\bo}{\mathbf{o}}
\newcommand{\bq}{\mathbf{q}}
\newcommand{\bu}{\mathbf{u}}
\newcommand{\bW}{\mathbf{W}}
\newcommand{\bR}{\mathbf{R}}
\newcommand{\bT}{\mathbf{T}}
\newcommand{\bz}{\mathbf{z}}
\newcommand{\bzast}{\bz^{\ast}}

\newcommand{\bomega}{\boldsymbol{\omega}}
\newcommand{\bOmega}{\boldsymbol{\Omega}}

\newcommand{\hbh}{\hat{\bh}}
\newcommand{\apara}{a_{\|}}

\newcommand{\baperp}{\mathbf{a}_{\perp}}

\newcommand{\bcL}{\boldsymbol{\mathcal{L}}}
\newcommand{\tauB}{\tau_\mathrm{D}}
\newcommand{\tauK}{\tau_\mathrm{M}}
\newcommand{\tauKast}{\tauK^{\ast}}
\newcommand{\tauI}{\tau_{I}}
\newcommand{\dd}{\text{d}}

\newcommand{\bone}{\mathbf{I}}
\newcommand{\LL}{\texttt{L}}
\newcommand{\feq}{f_\mathrm{eq}}
\newcommand{\freq}{\omega}
\newcommand{\zeq}{z_\mathrm{eq}}
\newcommand{\kb}{k_\mathrm{B}}
\newcommand{\etas}{\eta_\mathrm{s}}
\newcommand{\taueff}{\tilde{\tau}_\mathrm{eff}}
\newcommand{\ave}[1]{\langle #1 \rangle}
\newcommand{\aveeq}[1]{\langle #1 \rangle_{0}}

\begin{document}
\title{Magnetic susceptibility, nanorheology, and magnetoviscosity of magnetic nanoparticles in viscoelastic environments}

\author{Patrick Ilg, Apostolos E.A.S. Evangelopoulos}
\affiliation{School of Mathematical, Physical Sciences and Computational Sciences, University of Reading, Reading, RG6 6AX, UK}
\date{\today}

\begin{abstract}
While magnetic nanoparticles suspended in Newtonian solvents (ferrofluids) have been intensively studied in recent years, 
the effects of viscoelasticity of the surrounding medium on the nanoparticle dynamics are much less understood. 
Here we investigate a mesoscopic model for the orientational dynamics of isolated magnetic nanoparticles 
subject to external fields, viscous and viscoelastic friction as well as the corresponding random 
torques. 
We solve the model analytically in the overdamped limit for weak viscoelasticity. 
By comparison to Brownian Dynamics simulations we establish the limits of validity of the analytical solution. 
We find that viscoelasticity does not only slow down the magnetization relaxation, 
shift the peak of the imaginary magnetic susceptibility $\chi''$ to lower frequencies and 
increase the magnetoviscosity, it also leads to non-exponential relaxation and a broadening of $\chi''$. 
The model we study also allows to test a recent proposal for using magnetic susceptibility measurements 
as a nanorheological tool using a variant of the Germant-DiMarzio-Bishop relation. 
We find for the present model and certain parameter ranges  
that the relation of the magnetic susceptibility to the shear modulus is satisfied to a good approximation. 
\end{abstract}

\pacs{75.75.Jn, 75.50.Mn, 83.80.Hj, 05.40.Jc}

\maketitle

\section{Introduction}

Magnetic nanoparticles suspended in Newtonian solvents are known as ferrofluids \cite{rosensweigbook}. 
Since their conception in the 1960s, they have attracted considerable attention, due to their interesting properties, most notably the magnetoviscous effect \cite{KrogerIlgHess_JPHYS03}, and the breadth of applications \cite{Torres-Diaz2014}.
Ferrofluids are reasonably well understood today, at least for not too strong interactions \cite{Holm_review2005,Ilg_lnp}. 
More recently, new magnetic materials like soft magnetic elastomers and ferrogels have been synthesized 
\cite{Zrinyi_review,Annette_magnetlinkedgel,Tschoepe_ferrogel2011,Reddy_ferrogel,pi_hydrogel}. 
These new materials combine properties of both traditional elastomers and ferrofluids, because the suspending medium is neither a purely elastic solid nor a Newtonian liquid. 
Magnetic nanoparticles are also used more and more often in a number of biomedical applications such as drug delivery and magnetic hyperthermia 
\cite{Hergt2006,Alexiou2007}, magnetorelaxometry \cite{Wiekhorst:2012fz}, and biosensors \cite{Haun:2010kg}.
In these instances, magnetic nanoparticles are embedded in more complex, non-Newtonian environments. 
Magnetorelaxometry is an example of a powerful biomedical application, where analyzing the rotational diffusion of magnetic nanoparticles is used as a diagnostic tool \cite{Wiekhorst:2012fz}. 
Another example is the rotational microrheology of magnetic endosomes that has been employed to study the local viscoelasticity in cells 
\cite{Wilhelm2003}. 
However, for an accurate interpretation of all these experimental data a detailed knowledge of the magnetization dynamics 
in a complex environment is needed. 

The need for a better understanding of viscoelastic effects on the dynamics of magnetic colloids has prompted a number of experimental and theoretical investigations. 
Barrera \emph{et al.}~\cite{Barrera:2010kf} were one of the first to measure the dynamic susceptibility of magnetic nanoparticles suspended in fluids of varying viscoelasticity, which was tuned via progressive gelation of a gelatin solution. 
Recent experiments along similar lines have been performed and analyzed in terms of different, mostly phenomenological models 
\cite{Remmer:2017gf}. 
While these models are found to fit the experimental results on the imaginary part of the dynamic susceptibility reasonably well, 
they lead to different conclusions about an effective shear modulus. 
Similarly, experimental results on magneto-optical transmission on Nickel-nanorods suspended in different viscoelastic liquids 
were interpreted within Kelvin-Voigt and Maxwell models, which, however, resulted in calculated shear moduli different from the macroscopic measured ones \cite{Tschope:2014kj}. 
A careful study on dynamic susceptibility measurements has shown that for various polymeric solvents 
a variant of the Germant-DiMarzio-Bishop model can be used to infer the complex shear modulus \cite{Roeben:2014co}. 
Such nanorheological measurements are of great practical use as well as theoretical interest. 
However, their theoretical foundation in the present context remains unclear.

Despite a number of recent efforts, the dynamics of magnetic nanocolloids in viscoelastic environments remains  
considerably less well-studied than ferrofluids. 
Macroscopic approachs in terms of nonequilibrium thermodynamics established the hydrodynamic equation of ferrogels 
\cite{Pleiner_ferrogel,Attaran2017}. 
Some first steps in coarse graining a molecular model of polymer chains permanently attached to magnetic particles 
are presented in \cite{Pessot2015}. 
Mesoscopic models for the rotational dynamics of single-domain magnetic nanocolloids  
have been investigated for some time (see \cite{Volkov:2001jr,Raikher2005,Raikher_MVEwithGLE} and references therein), 
where viscoelasticity effects are modeled in terms of non-Markovian Langevin or Brownian dynamics.  
More recently, the need for modeling the viscoelastic medium by a generalized Maxwell model \cite{Raikher:2013ju} 
together with a proper three-dimensional treatment of rotations has been recognized \cite{Rusakov:2017gi}. 
 
Here, we study the same model for the three-dimensional rotational dynamics of single-domain magnetic nanoparticles in a 
generalized Maxwell fluid as proposed in Ref.~\cite{Rusakov:2017gi}. 
We solve the model analytically in case of weak viscoelasticity. Brownian dynamics simulations establish the range of validity 
of the analytical result and provide numerical results for the full range of viscoelasticity effects. 
Our results show that the Germant-DiMarzio-Bishop model works surprisingly well for certain parameter ranges, 
which provides a theoretical justification for the nanorheological measurements advocated in Ref.~\cite{Roeben:2014co}.

The paper is organized as follows. 
The model that we study here is presented in Sect.~\ref{model.sec}, including the overdamped limit as well as 
a description of its numerical solution via Brownian Dynamics simulations. 
The analytical as well as numerical results are presented in Sect.~\ref{results.sec}. 
Finally, some conclusions are offered in Sect.~\ref{concl.sec}. 

\section{Kinetic model of magnetic nanoparticles in viscous and viscoelastic media} \label{model.sec}

Consider an ensemble of statistically independent, non-interacting colloidal magnetic nanoparticles (i.e.~the ultra-dilute limit). 
Let the orientation of the particle be described by the three-dimensional unit vector $\bu$ ($\bu^2=1$). 
For given angular velocity $\bomega$, the orientational motion is given by $\dot{\bu}=\bomega\times\bu$.
The equation of motion is given by the angular momentum balance
$ I\dot{\bomega} = \bT$, 
where $I$ is the moment of inertia of the particle 
and $\bT$ the sum of all torques acting on the particle.

Here, we want to consider torques exerted by an external magnetic field $\bH$ as well as torques exerted by the surrounding viscous and viscoelastic medium. 
In particular, we make the following assumptions: 
(i) rigid dipole approximation; i.e.~the dipole moment remains always parallel to the particle orientation, 
${\bf m}=m \bu$ where $m$ denotes the magnetic moment of a single colloidal particle. 
With this assumption we neglect internal N\'eel relaxation, so we are restricted to large enough particles and not too high frequencies 
\cite{rosensweigbook}. 
Under this assumption, the torque due to an external magnetic field $\bH$ is given by 
$\bT^H=-\bcL \Phi = m\bu\times\bH$, where $\Phi=-m\bu\cdot\bH$ is the Zeeman potential energy and 
$\bcL=\bu\times\partial/\partial\bu$ the rotational operator. 
We also assume (ii) that we can model the torque due to the 
viscous solvent by rotational friction $\xi=8\pi\etas a^{3}$ and corresponding white noise with thermal energy $\kb T$, where $\etas$ is the solvent viscosity and $a$ the hydrodynamic radius of the particle. 
(iii) Furthermore, we assume that we can model the viscoelastic contribution by the retarded friction $\zeta(t)$ and corresponding random torque $\bR$. 
(iv) Finally, we assume dilute conditions such that inter-particle interactions are negligible. 
 Also effects due to hydrodynamic memory are assumed to be irrelevant, since we will later consider long enough time scales. 

Under these assumptions we arrive at the generalized Langevin equation for the rotational motion
\begin{equation} \label{GLE}
 I\dot{\bomega}(t) = m\bu(t)\times\bH(t) - \xi[\bomega(t) - \bOmega(t)]
 -  \int_0^t\!\dd t'\, \zeta(t-t')[\bomega(t')-\bOmega(t')]  + \sqrt{2\kb T\xi}\,\dot{\bW}_{t}
 +\bR(t).
\end{equation}
The torque due to friction is only experienced if the particle's angular velocity $\bomega$ differs from the angular velocity $\bOmega$ 
of the surrounding medium. 
White noise associated with the viscous solvent is modeled by the three-dimensional Wiener process $\bW_{t}$. 
The viscoelastic contribution exerts not only a friction torque described by the retarded rotational friction $\zeta(t)$ but also 
a random torque $\bR$ with $\ave{\bR}=0$. 
The fluctuation-dissipation theorem (FDT) requires that these torques are related by 
$\ave{\bR(t)\bR(t')}=\kb T \bone\zeta(t-t')$ with $\bone$ the three-dimensional unit matrix \cite{Mazenko}.  

In the following, we consider the viscoelastic contribution as a single-mode Maxwell model where 
the memory kernel can be expressed as 
$\zeta(t)=(\zeta_{0}/\tauK)e^{-t/\tauK}$, where $\tauK$ is the relaxation time 
and $\zeta_{0}$ the rotational friction coefficient of the particle in the viscoelastic medium. 
In this case, we conclude from FDT that $\bR$ is exponentially correlated, i.e.~a three-dimensional Ornstein-Uhlenbeck process, 
 $\ave{\bR(t)\bR(t')}=k_{\rm B}T\zeta_{0}/\tauK \exp{[-(t-t')/\tauK]}\bone$, i.e.~$\bR$ obeys the stochastic differential equation 
 \begin{equation} \label{OU}
\dd \bR(t) = - \frac{1}{\tauK}\bR(t){\rm d}t +  B_{R}{\rm d}\bW_{t}^{R},
 \end{equation}
 with $B_{R}=\sqrt{2kT\zeta_{0}}/\tauK$. 
 Here, $\bW_{t}^{R}$ is another three-dimensional Wiener process, statistically independent of $\bW_t$. 
 In the limit $\tauK\to 0$, we obtain $\zeta(t)=\zeta_{0}\delta(t)$, i.e.~a Newtonian bath 
 and the system reduces to the corresponding  model of ferrofluids \cite{Ilg_lnp} with an effective rotational friction coefficient 
 $\xi+\zeta_{0}$. 
 The Maxwell model has frequently been employed to model the effect of viscoelastic media on colloid dynamics \cite{Raikher2005}. 
For polymer solutions, however, it has been argued that a generalized Maxwell (also termed Jeffrey's) model that we employ here 
is more appropriate 
to properly describe the high-frequency behavior \cite{Grimm:2011ei,Raikher:2013ju}.

\subsection{Extended variable formalism}
The stochastic integro-differential equation (\ref{GLE}) with (\ref{OU}) can be converted into a system of stochastic differential equations 
by introducing the auxiliary variable 
\begin{equation}
\bz(t) = -\int_{0}^{t}\!\dd t' \zeta(t-t')[\bomega(t')-\bOmega(t')] + \bR(t),
\end{equation}
which can be interpreted as the effective torque due to friction and noise of the viscoelastic environment. 

For the exponential memory kernel, the time derivative is particularly simple and we arrive at the following 
system of stochastic differential equations
\begin{align}
\dd \bu & = \dd \bomega \times \bu \label{eq:du}\\
I\dd \bomega & = [-\bcL \Phi - \xi(\bomega-\bOmega) + \bz]\dd t + \sqrt{2\kb T\xi}\,\dd \bW \label{eq:domega}\\
\dd \bz & = -\frac{1}{\tauK} [ \bz + \zeta_{0}(\bomega-\bOmega)]\dd t + B_{R}\,\dd \bW^{R}. \label{eq:dz}
\end{align}

Define the probability density $F(\bu,\bomega,\bz;t)$. 
Then $F$ obeys the Fokker-Planck equation corresponding to the stochastic differential equations (\ref{eq:du}-\ref{eq:dz})
\begin{align} \label{FPE}
\frac{\partial}{\partial t}F = & -\bcL\cdot(\bomega F) 
- \frac{1}{I}\frac{\partial}{\partial\bomega}\cdot[(-\bcL\Phi - \xi(\bomega-\bOmega)+\bz)F] 
+ \frac{\kb T\xi}{I^{2}}\frac{\partial^{2}}{\partial\bomega^{2}} F \nonumber\\
&  - \frac{\partial}{\partial\bz}\cdot[(-\frac{1}{\tauK}\bz - \frac{\zeta_{0}}{\tauK}(\bomega-\bOmega))F] 
+ \frac{\kb T\zeta_{0}}{\tauK^{2}}\frac{\partial^{2}}{\partial\bz^{2}} F.
\end{align}

In the absence of flow, $\bOmega=0$, the stationary solution to Eq.~(\ref{FPE}) is given by the Boltzmann distribution 
\begin{equation} \label{F0}
F_{0}(\bu,\bomega,\bz) = Z^{-1}\exp{[-\beta\Phi(\bu) - \frac{1}{2}\beta I\bomega^{2} - \frac{\bz^{2}}{2\zeq^{2}}]},
\end{equation}
with normalization constant $Z$, $\beta=1/(\kb T)$, and $\zeq^{2}=\zeta_{0}/(\beta\tauK)$. 
The equilibrium probability density factorizes and therefore the equilibrium magnetization is given by the Langevin function and is independent of 
the (retarded) friction, as it should. 
Furthermore, the equilibrium Gaussian fluctuations of the angular velocity $\bomega$ and torques $\bz$ are specified by 
$\ave{\omega_{\alpha}\omega_{\beta}}_{0}=(\beta I)^{-1}\delta_{\alpha\beta}$ and 
$\ave{z_{\alpha} z_{\beta}}_{0} = \zeq^{2}\delta_{\alpha\beta}$.

\subsection{Overdamped limit}
By construction, the model presented above exhibits three relaxation times: 
the inertial relaxation time $\tauI=I/\xi$, 
the Brownian relaxation time of rotational diffusion due to the viscous solvent $\tauB = \xi/(2\kb T)$, 
and the relaxation time of the viscoelastic contribution $\tauK$. 

For colloidal particles in general and magnetic nanoparticles in particular, the usual condition 
$\tauI\ll \tauB,\tauK$ holds \cite{Dhont_book}. 
Therefore, we consider in the following the overdamped limit $\tauI\to 0$, i.e.~Eq.~(\ref{GLE}) for $I\dot{\bomega}\to 0$. 
Note, however, that measurements of the magnetic susceptibility have shown the influence of inertia effects at 
sufficiently high frequencies \cite{Fannin95}. 
In the overdamped limit, we can eliminate the angular velocity $\bomega$ from Eq.~(\ref{eq:domega}) as independent variable  
and arrive at the reduced set of stochastic differential equations 
\begin{align}
 \dd \bu & = [\bOmega - \frac{1}{\xi}\bcL \Phi + \frac{1}{\xi}\bz]\times\bu\,\dd t + B\dd \bW \times \bu \label{eq:du_2}\\
 \dd \bz & = [-\frac{1+q}{\tauK}\bz + \frac{q}{\tauK}\bcL \Phi] \dd t + B_{R}(\dd \bW^R - \sqrt{q}\dd \bW), \label{eq:dz_2}
\end{align}
where $B=\sqrt{2\kb T/\xi}$ and $q=\zeta_{0}/\xi$ the ratio of the friction coefficients. 
Note that taking the overdamped limit leads to the appearance of correlated noise in the auxiliary variable $\bz$. 
The Fokker-Planck equation for the reduced probability density 
$f(\bu,\bz;t)$ corresponding to Eqs.~(\ref{eq:du_2}) and (\ref{eq:dz_2}) reads \cite{hcobook} 
\begin{align} 
 \frac{\partial}{\partial t} f = & -\bcL \cdot [(\bOmega - \frac{1}{\xi}\bcL \Phi + \frac{1}{\xi}\bz)f] + \frac{kT}{\xi}\bcL^2 f \nonumber\\
  & -\frac{\partial}{\partial \bz}\cdot[ (-\frac{1+q}{\tauK}\bz + \frac{q}{\tauK}\bcL \Phi) f] 
  + \frac{\kb T\zeta_{0}}{\tauK^{2}}(1+q) \frac{\partial^{2}}{\partial \bz^{2}} f 
   - q\frac{2\kb T}{\tauK} \frac{\partial}{\partial{\bz}}\cdot\bcL f .\label{FPE_uz}
\end{align}
Equation (\ref{FPE_uz}) agrees with the corresponding Fokker-Planck equation in \cite{Rusakov:2017gi}. 
We note that in the limit $q\to 0$, we recover the kinetic model for (non-interacting) ferrofluids \cite{Ilg_lnp}. 
In the absence of flow, $\bOmega=0$, we find the equilibrium Boltzmann distribution 
\begin{equation} \label{feq-uz}
f_{0}(\bu,\bz) = Z_{1}^{-1}\exp{[-\beta\Phi(\bu) - \bz^{2}/(2\zeq^{2})]}
\end{equation}
as stationary solution to Eq.~(\ref{FPE_uz}). 
Note that $f_{0}$ can also be obtained from $F_{0}$, Eq.~(\ref{F0}), when integrated over $\bomega$, as it should be 
and $Z_{1}$ is the corresponding normalization constant. 
We emphasize that stationary properties should not be affected by taking the overdamped limit. 
However, this is not the necessarily the case in some earlier works where only a viscoelastic bath is present and additional manipulations need to be invoked \cite{Raikher2005}. 

For later use, we introduce the Fokker-Planck operator $\LL$ from Eq.~(\ref{FPE_uz}) via 
$\partial f/\partial t = \LL f$. Separating the effects of internal dynamics $\LL_{0}$, the external field $\LL_{h}$ and external flow $\LL_{\Omega}$ allows the decomposition 
$\LL=\LL_{0} + \LL_{h} + \LL_{\Omega}$. 
Furthermore, we define the adjoint operators $\LL^{\dagger}$ by 
$\int\!\dd\bu\dd\bz A \LL f  = \int \!\dd\bu\dd\bz f \LL^{\dagger} A $ for an arbitrary function $A=A(\bu,\bz)$. 
Using integration by parts we derive from (\ref{FPE_uz}) the explicit form of the adjoint operators, 
\begin{align}
 \tauB \LL_{0}^{\dagger} A = &  \frac{1}{2}\bcL^2 A 
 -\frac{1+q}{\tauKast}\bzast\cdot\frac{\partial}{\partial{\bzast }}A + \epsilon\bzast\cdot\bcL A \nonumber\\
 & + \frac{1+q}{\tauKast}\frac{\partial^{2}}{\partial{\bzast}^{2}} A - 2\epsilon\frac{\partial}{\partial{\bzast}}\cdot\bcL A  \label{L0def}\\
 \tauB \LL_{h}^{\dagger}(t) A = & (\bu\times\bh(t))\cdot[\frac{1}{2}\bcL A - \epsilon \frac{\partial}{\partial{\bzast}}A]\label{Lhdef}\\
 \tauB \LL_{\Omega}^{\dagger}(t) A = & \bOmega^{\ast}(t)\cdot\bcL A,
\end{align}
where we introduced the dimensionless quantities $\bzast=\bz/\zeq, \bOmega^{\ast}=\tauB\bOmega$, 
the ratio of relaxation times $\tauKast=\tauK/\tauB$, 
$\bh=m\bH/(\kb T)$ 
and 
$\epsilon = \sqrt{q/(2\tauKast)} = \tauB \zeq/\xi$.
Note that the parameter $\beta=4\epsilon^{2}$ was used in Ref.~\cite{Rusakov:2017gi} and interpreted as ``springiness''. 

\subsection{Brownian Dynamics simulations}
For comparison to analytical results, we also perform Brownian dynamics simulations of the model equations. 
Thereby we avoid certain assumptions detailed below that we employ for the analytical calculations. 

We have implemented the stochastic differential equations (\ref{eq:du_2}) and (\ref{eq:dz_2}) 
using a first-order Euler scheme as well as a second order Heun scheme \cite{hcobook}. 
For the current purpose, both schemes give identical results for small enough time steps. 
For simplicity, we here only describe the Euler scheme. 

Using the dimensionless quantities introduced above with $t^{\ast}=t/\tauB$, 
we define the dimensionless increment of the angular velocity $\Delta\bomega^{\ast}=\tauB\Delta\bomega$ by 
\begin{equation}
\Delta\bomega^{\ast} = (-\frac{1}{2}\bu\times\bh + \epsilon\bzast)\Delta t^{\ast} + \Delta\bW^{\ast}
\end{equation}
where $\Delta\bW^{\ast}$ denotes a dimensionless three-dimensional Wiener increment, i.e.~a three-dimensional Gaussian random variable with zero mean and variance $\Delta t^{\ast}$. 
With the help of $\Delta\bomega^{\ast}$, the increments of the variables $\bu$ and $\bzast$ over a short time step $\Delta t^{\ast}$ are given by 
\begin{align}
\Delta\bu & = (\bOmega^{\ast}\Delta t^{\ast} + \Delta\bomega^{\ast}) \times \bu \label{BDu}\\
\Delta\bzast & = - \frac{1}{\tauKast}\bzast\Delta t^{\ast} - 2\epsilon\Delta\bomega^{\ast} + \sqrt{2/\tauKast}\Delta\bW^{\ast\ast} \label{BDz}
\end{align}
where $\Delta\bW^{\ast\ast}$ is another three-dimensional Gaussian random variable with zero mean and variance $\Delta t^{\ast}$, statistically independent of $\Delta\bW^{\ast}$. 

We found $\Delta t^{\ast} = 10^{-3}$ to be small enough to obtain identical results to the corresponding Heun scheme for the observables of interest. An ensemble of $N=5\times10^{5}$ independent realizations of $\bu$ and $\bzast$ was usually used 
in order to obtain reliable estimates for mean values. 

\section{Results} \label{results.sec}

\subsection{Short-time rotational diffusion}
Since the orientation $\bu$ is restricted to the three-dimensional unit sphere, rotational diffusion is defined for short times only from the relation 
\begin{equation} \label{Ddef}
\aveeq{(\bu(t)-\bu(0))^2} = 4Dt, \quad t\ll 1/(2D)
\end{equation}
with the rotational diffusion coefficient $D$.  
Averages $\ave{\bullet}_0$ are taken with respect to the equilibrium initial ensemble $f_{0}$, Eq.~(\ref{feq-uz}). 
In terms of the auto-correlation function $C(t)=\aveeq{\bu(t)\cdot\bu(0)}$, the mean-squared orientational displacement 
can be expressed as 
$\aveeq{(\bu(t)-\bu(0))^2}=2(1-C(t))$. 
The short-time behavior of $C(t)$ can be computed from the Taylor series 
$C(t) = 1 + \dot{C}(0)t + \frac{1}{2}\ddot{C}(0)t^{2} + {\cal O}(t^{3})$, 
where we used the fact that $C(0)=\aveeq{\bu^{2}} =1$. 
The first and second order terms are calculated from $\dot{C}(0)=\aveeq{[\LL^{\dagger}\bu]\cdot\bu}$ and 
$\ddot{C}(0)=\aveeq{[(\LL^{\dagger})^{2}\bu]\cdot\bu}$, respectively. 
Some details of the calculation can be found in appendix \ref{app:Cshort}. 
In the absence of external flow we find 
\begin{equation} \label{uucorrel}
 \aveeq{(\bu(t)-\bu(0))^2} = \frac{2t}{\tauB} - \left(1 + 2\epsilon^{2} + \frac{1}{2}hL_{1}(h)\right)\left(\frac{t}{\tauB}\right)^2 + \mathcal{O}(t^3), 
\end{equation}
where $L_{1}(h)=\coth(h)-1/h$ denotes the Langevin function. 
Interestingly, the first order term is independent of the viscoelastic bath and external magnetic field 
that both contribute only from the second order on. 
Thus, in view of the definition (\ref{Ddef}), we conclude that $D=1/(2\tauB)$ 
is identical to the viscous diffusion on short time scales $t\ll\tauB$.

\subsection{Magnetization relaxation}
In magnetorelaxometry, the relaxation of the magnetization is analyzed after a strong ordering field is switched off 
\cite{Wiekhorst:2012fz}. 
The present treatment does not include internal N\'eel relaxation that is important for a proper interpretation of 
the corresponding experimental results. Nevertheless, we here provide a detailed analysis of the effect of medium viscoelasticity on the 
magnetization relaxation for magnetically hard nanoparticles. 
Our results would therefore be useful for separating out viscoelasticity effects due to the biological environment 
from the magnetization relaxation signal. 

Assume that $\bOmega=0$ and a strong ordering magnetic field has been applied along the $z$-direction for a sufficiently long time so that the Boltzmann equilibrium (\ref{feq-uz}) is established. 
At time $t=0$, the external field is switched off instantaneously.  
Then, the reduced magnetization $M/M_\text{sat}=\ave{u_{z}}$ obeys the ordinary differential equation 
$\frac{\dd}{\dd t}\ave{u_{z}}=\ave{\LL_{0}^{\dagger}u_{z}}$ with 
initial condition $\ave{u_{z}}(0)=1$. 
Define $a_{0} = \ave{u_{z}}, a_{1}=\ave{o_{z}}, a_{2}=\ave{q_{z}}$
with $o_{\alpha}=-\epsilon_{\alpha\beta\gamma}u_{\beta}z^{\ast}_{\gamma}$, 
$q_{\alpha} = \epsilon_{\alpha\beta\gamma}\epsilon_{\mu\beta\lambda}u_{\lambda}z^{\ast}_{\gamma}z^{\ast}_{\mu}$. 
With these quantities, the first members of the moment hierarchy read 
\begin{align}
\tauB\dot{a}_{0} & = - a_{0} + \epsilon a_{1} \label{dta0}\\
\tauB\dot{a}_{1} & = -\lambda_{1} a_{1} + \epsilon a_{2} + 4\epsilon a_{0} \label{dta1}\\
\tauB\dot{a}_{2} & = -\lambda_{2} a_{2} + \epsilon a_{3} + 6\epsilon a_{1} - \frac{4}{\tauKast} a_{0} 
- 8\epsilon^{2} a_{0}, \label{dta2}
\end{align}
where $\lambda_{n}=1+n/\tauKast+2n\epsilon^{2}$. 
Note that only in the limit $\epsilon\to 0$, i.e.~in the absence of viscoelastic contributions, the moment system truncates 
at the first order. In general, we are faced with an infinite hierarchy for which an exact solution is unknown. 

To make further progress analytically, we look for solutions as power series in $\epsilon$, 
\begin{equation} \label{series-epsilon}
a_{k}(t) = \sum_{n=0}^{\infty} \epsilon^{n} a_{k}^{(n)}(t).
\end{equation}
Inserting the expansion (\ref{series-epsilon}) into Eqs.~(\ref{dta0})-(\ref{dta2}) and matching equal orders of $\epsilon$ we find for ${\cal O}(\epsilon^{0})$ a single-exponential decay, 
$a_{0}^{(0)}(t) =  e^{-t/\tauB}$, corresponding to the purely viscous limit of the model. 
Matching also next orders in $\epsilon$ we find (see Appendix \ref{append:Mrelax} for some details of the calculations) 
 \begin{equation} \label{Mrelax}
\ave{u_{z}}(t) = e^{-t/\tauB} \left\{ 1 +  q 
\left[ 
\frac{t}{\tauB} + \tauKast \left( -1 + e^{-t/\tauK} \right)
\right]
\right\} + {\cal O}(\epsilon^{3}).
 \end{equation}

\begin{figure}[htb]
 \includegraphics[width=0.47\textwidth]{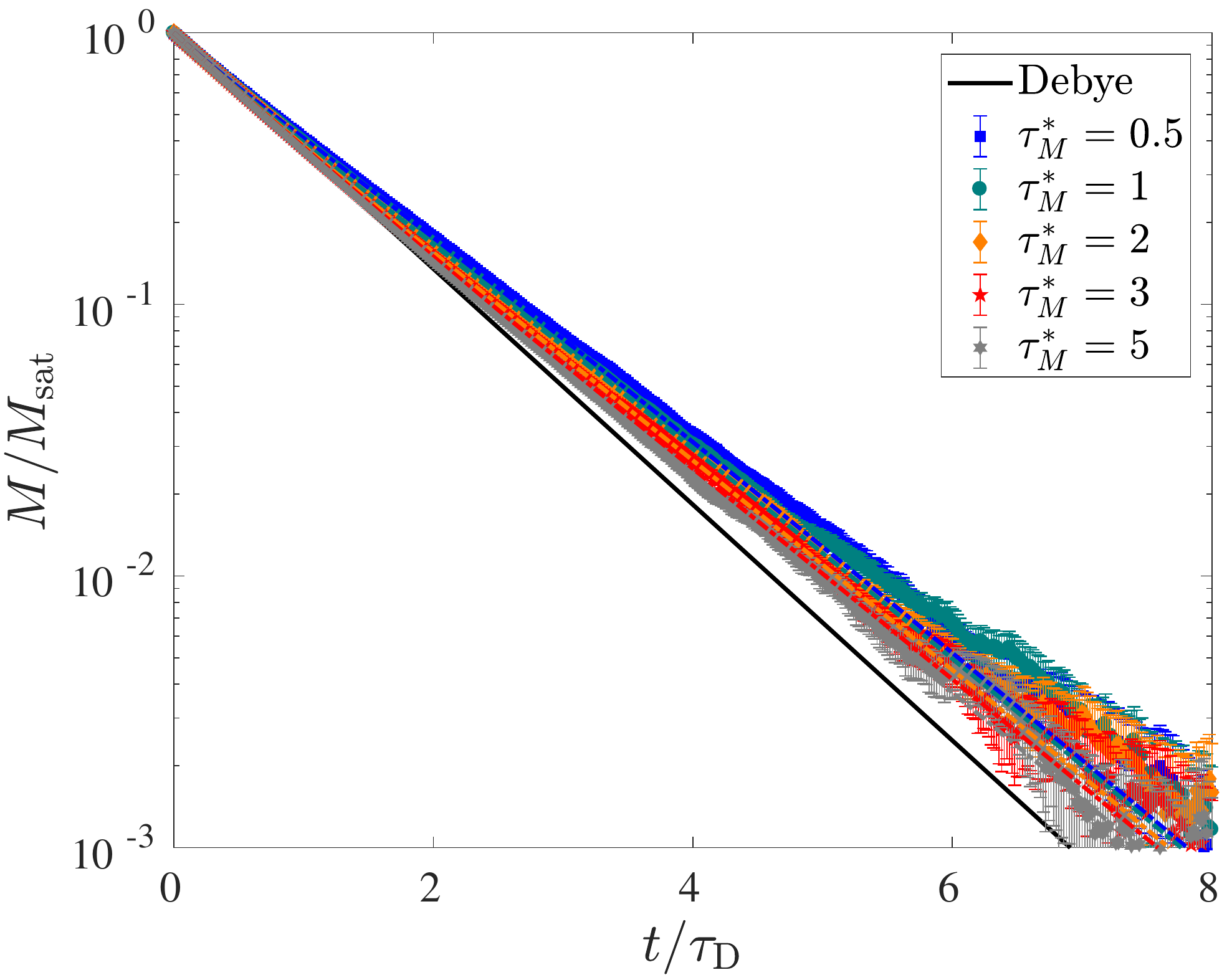}
 \caption{Magnetization relaxation $M(t)/M_{\rm sat}=\ave{u_z}(t)$ after switching off a strong 
 ordering field. Parameters are chosen as $q=0.2$ and $\tauKast$ varying from $\tauKast=0.5$ to $5$ as indicated 
 in the legend. 
 Symbols and solid lines correspond to simulation and analytical results from Eq.~(\ref{Mrelax}), respectively. 
 For comparison, the solid line is the Debye law $\exp{[-t/\tauB]}$.}
 \label{Mrelax.fig}
\end{figure}

\begin{figure}[htb]
 \includegraphics[width=0.47\textwidth]{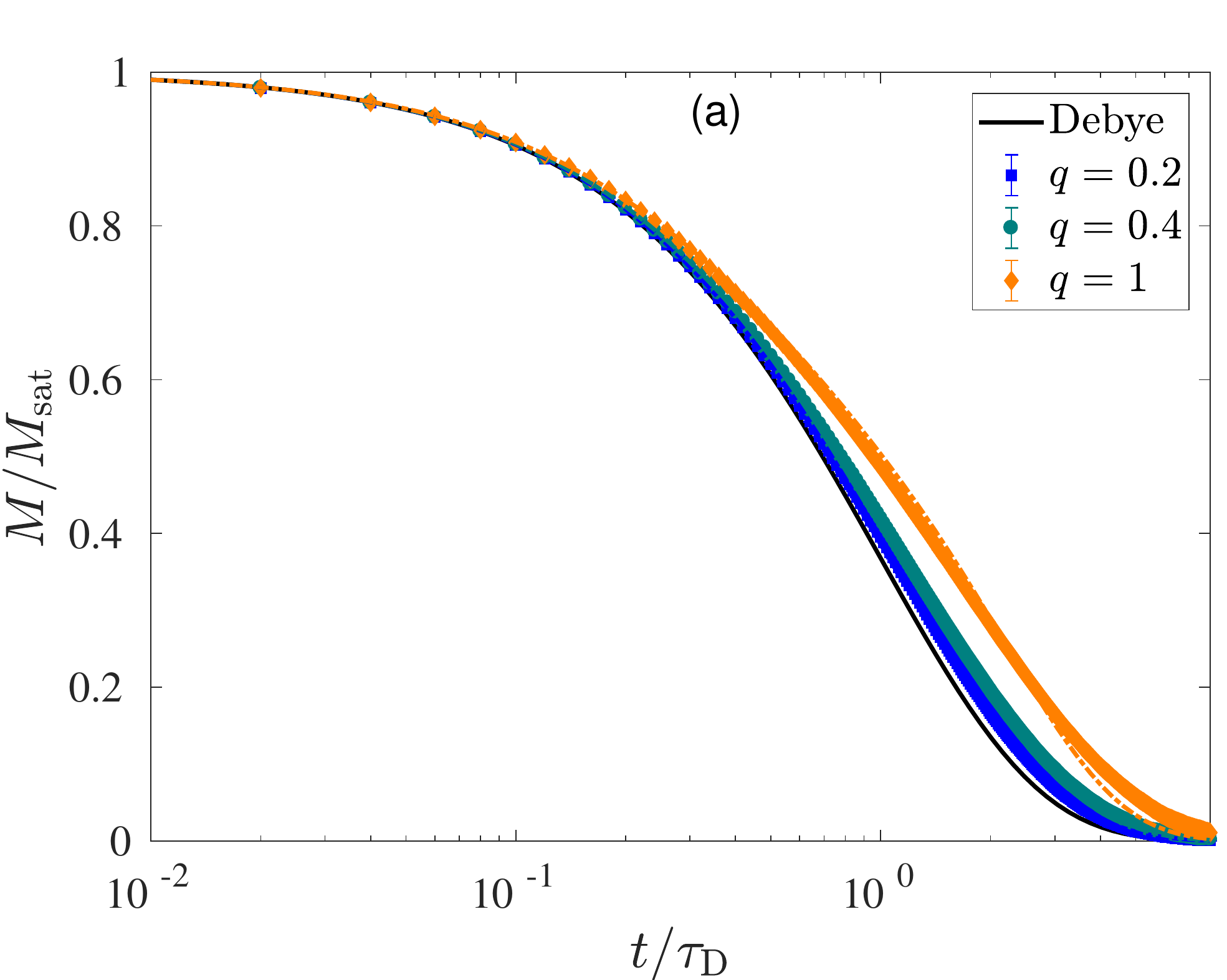}
 \includegraphics[width=0.47\textwidth]{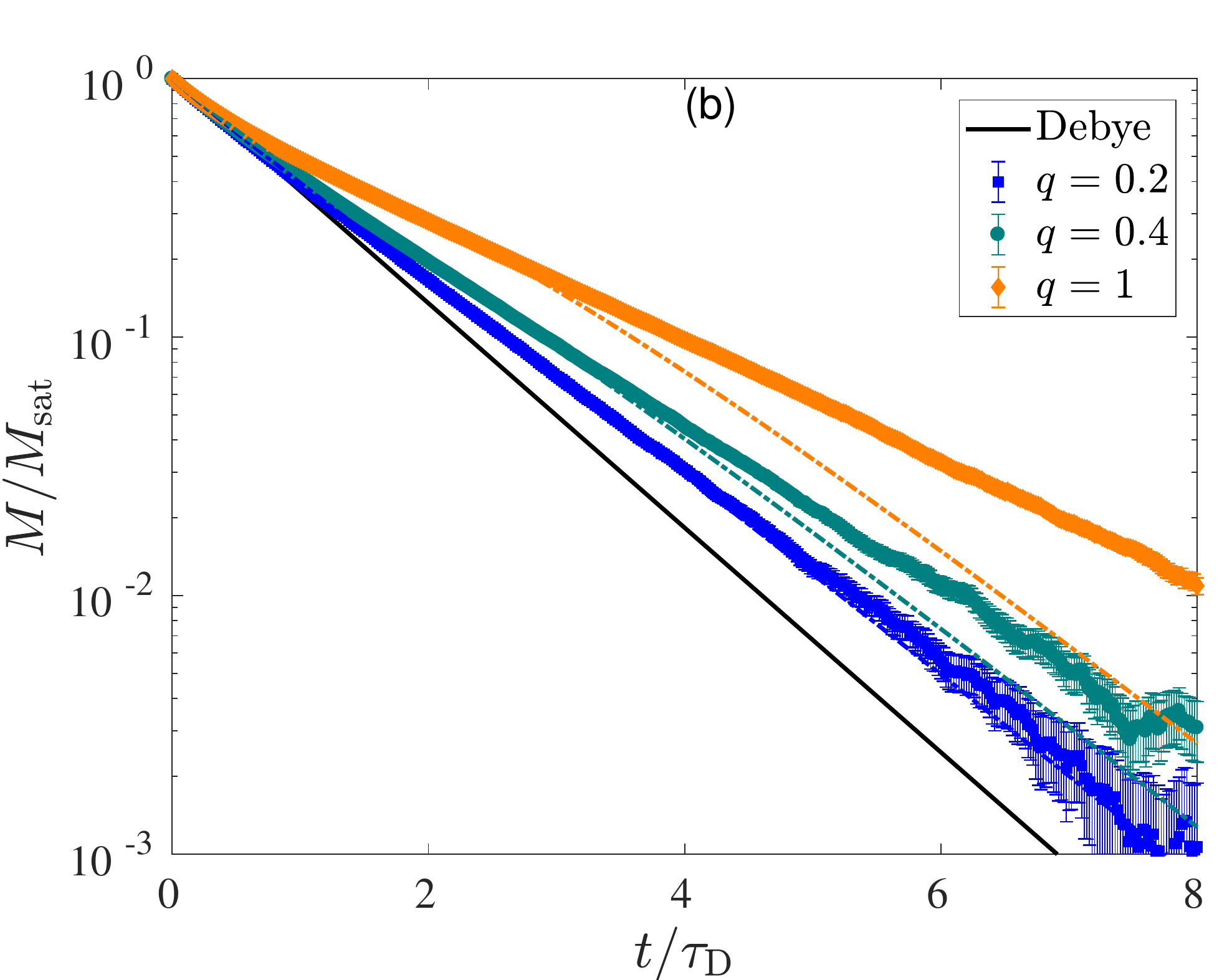}
 \caption{Magnetization relaxation $M(t)/M_{\rm sat}=\ave{u_z}(t)$ as in Fig.~\ref{Mrelax.fig} 
 but for parameters $\tauKast=1$ and $q=0.2, 0.4, 1$ from bottom to top.
 Panel (a) and (b) show the same data but on different axis scales.}
 \label{Mrelax2.fig}
\end{figure}

It is interesting to note that the magnetization relaxation from Eq.~(\ref{Mrelax}) is not simply given by a superposition of 
two exponentials. In other words, the viscous and viscoelastic contributions to the relaxation can not be considered independent. 
Figures \ref{Mrelax.fig} and \ref{Mrelax2.fig} show the magnetization relaxation $\ave{u_{z}}(t)$ on a semi-logarithmic scale. 
The analytical formula (\ref{Mrelax}) is compared with results from Brownian dynamics simulations of the model. 
Deviations from the single-exponential behavior of the purely viscous model are obvious. 
Increasing viscoelastic contributions slows down the magnetization relaxation more and more. 
The approximate formula (\ref{Mrelax}) provides an accurate description for the whole relaxation process when 
$q\lesssim 0.2$, whereas for larger values of $q$ only the early stages of the relaxation are captured correctly by Eq.~(\ref{Mrelax}) while for late stages the magnetization is underestimated. 

\subsection{Magnetic susceptibility}
We are interested in the linear response of the system to an externally applied weak magnetic field $\bH(t)$. 
When the dimensionless magnetic field $\bh(t)=m\bH(t)/(\kb T)$ is small, a first order perturbation expansion gives 
\begin{equation} \label{linresponse-h}
 f(t) =  \feq + \frac{1}{\tauB}\int_0^t\!\dd t'\, e^{(t-t')\LL_0} (\bu + \frac{\tauB}{\xi}\bu\times\bz)\cdot\bh(t') \feq  + {\cal O}(h^2)
\end{equation}
where we assumed that the system was initially in equilibrium, $f(0)=\feq$, i.e.~by Eq.~(\ref{feq-uz}) for $\Phi=0$. 
For simplicity of notation, we here suppress the arguments $(\bu,\bz)$ of the probability density $f$ and $\feq$. 
The induced magnetization $\bM(t)=nm\int \!\dd\bu\dd\bz\, \bu f(t)$ is therefore linearly related to the applied field when $\bh$ is small enough, 
$M_{\alpha}(t)=\int_{0}^{t}\!\dd t' \chi_{\alpha\beta}(t-t')H_{\beta}(t')$ with 
$ \chi_{\alpha\beta}(t) = \frac{nm^2}{kT\tauB} \ave{u_\alpha\, e^{t\LL_0} (\bu + \frac{\tau}{\xi}\bu\times\bz)_\beta }_{\rm eq}$. 
With the help of the adjoint operator $\LL_0^\dagger$ defined in Eq.~(\ref{L0def}) and 
with $\dot{\bu}=\LL_0^\dagger \bu$, the susceptibility tensor can be written as 
$\chi_{\alpha\beta}(t) = - \frac{nm^2}{kT\tauB} \frac{\dd}{\dd t} \ave{u_\alpha(t) u_\beta(0) }_{\rm eq}$. 
Since the system is isotropic in the absence of an external field, $\chi_{\alpha\beta}(t)=\chi(t)\delta_{\alpha\beta}$.
With the one-sided Fourier-transform, 
$ \tilde{\chi}(\freq) = \int_0^\infty\!\dd t'\, \chi(t) e^{-i\freq t}$, 
the frequency-dependent complex susceptibility takes the usual form \cite{Mazenko}
\begin{equation} \label{chi-omega}
\tilde{\chi}(\freq) = \chi_0 - \chi_{0} i\freq \int_0^\infty\!\dd t\, \ave{\bu(t)\cdot\bu(0)}_{\rm eq} e^{-i\freq t} 
\end{equation}
where $\chi_0=nm^2/(3\kb T)$ is the static (zero-frequency) susceptibility.

\begin{figure}[htb]
 \includegraphics[width=0.49\textwidth]{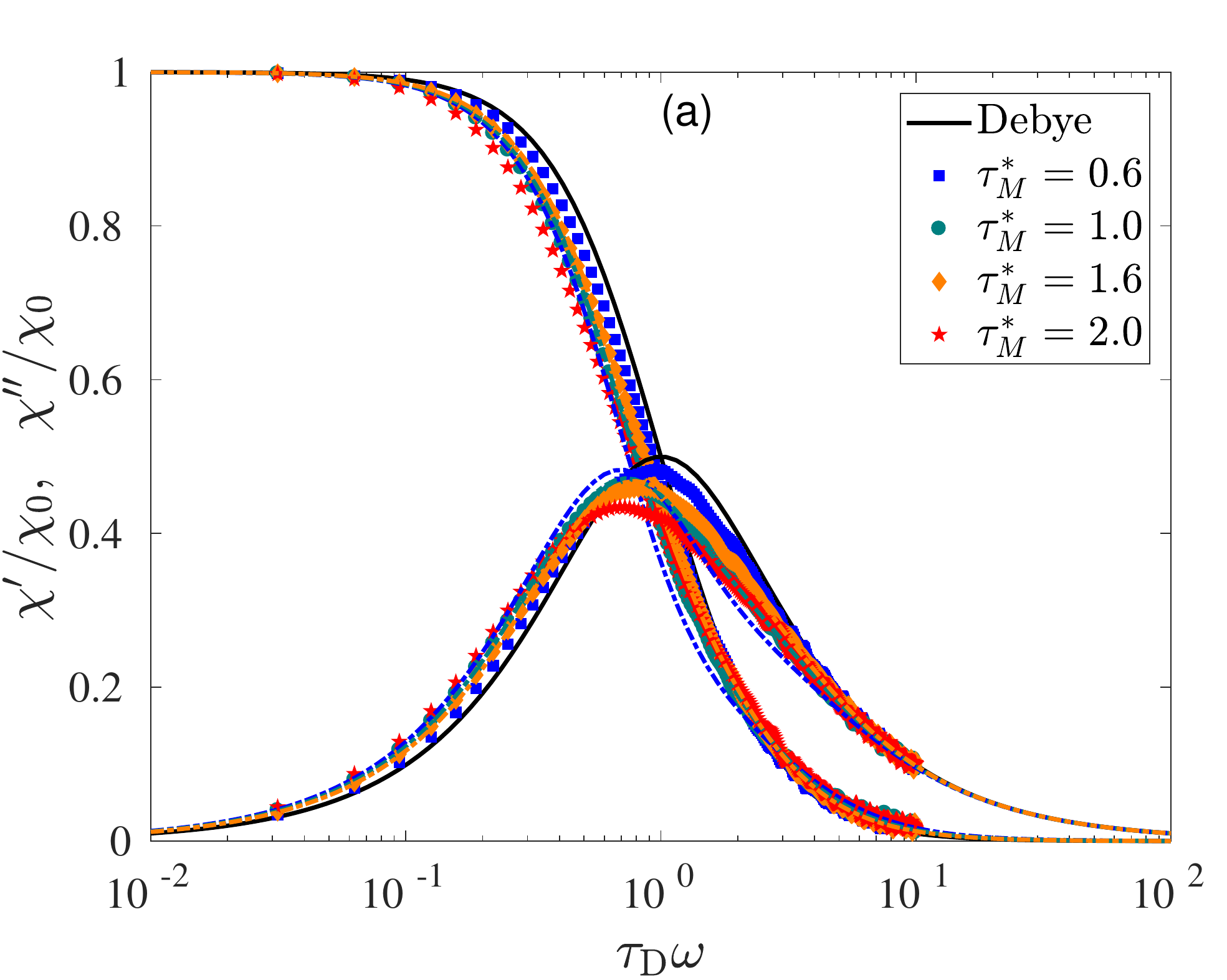}
 \includegraphics[width=0.49\textwidth]{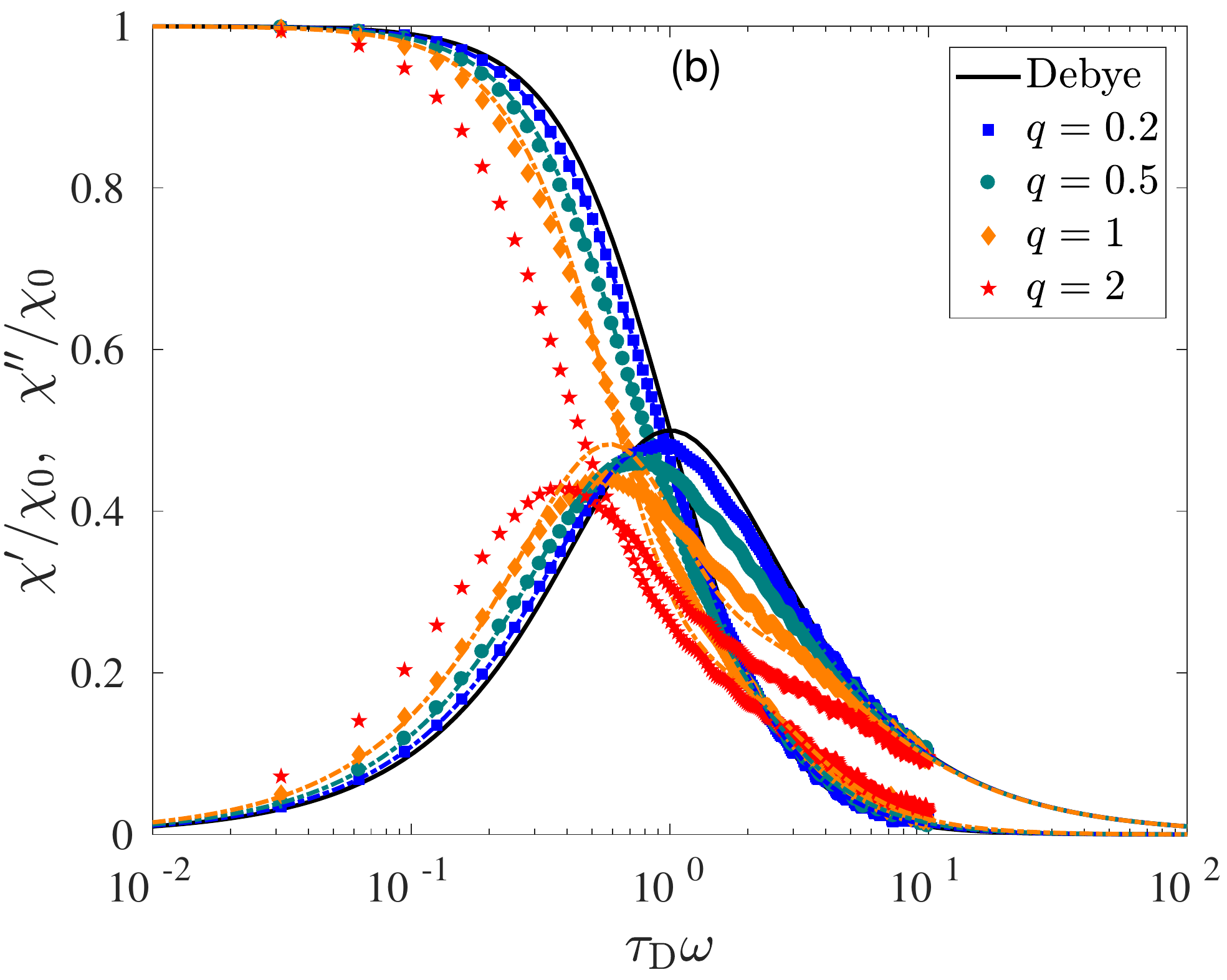}
 \caption{Real and imaginary part of the complex susceptibility as a function of dimensionless frequency 
 $\tauB\omega$ of the applied magnetic field. 
 Left panel shows results for $q=0.5$ and different values of $\tauK$, whereas in the right panel the value 
 $\tauKast=1.0$ was fixed and different values for the parameter $q$ were chosen. 
 Symbols and dashed lines correspond to simulation and analytical results from Eqs.~(\ref{chiprime}, \ref{chidprime}), respectively. 
 For comparison, the solid black line shows the Debye susceptibility.}
 \label{chi.fig}
\end{figure}

For small $q$, we  find that $C(t)=\ave{\bu(t)\cdot\bu(0)}_{\rm eq}$ obeys the same differential equation 
(\ref{dta0})-(\ref{dta2}) 
with the same initial condition and therefore $C(t)$ is also given by Eq (\ref{Mrelax}). 
Note that this is a special case of the general fluctuation-dissipation relation between relaxation and correlation functions \cite{Mazenko}. 
From the Laplace transform $\tilde{C}(s)=\int_0^\infty\!\dd t\, C(t)e^{-st}$  
the complex susceptibility can be obtained via $\tilde{\chi}(\freq) = \chi_0[1 - i\freq\tilde{C}(i\freq)]$ as 
\begin{align}
 \tilde{\chi}(\freq)/\chi_0 
  & =  \frac{1 + i q \tauK\freq}{1+i\tauB \freq} 
 -  \frac{i q \tauKast \tau_{1}\freq}{1+ i \tau_1 \freq}
 - \frac{iq\tauB\freq}{(1+i\tauB \freq)^2}
\end{align}
where $1/\tau_1=1/\tauB+1/\tauK$ is the effective relaxation time of the combined viscous and viscoelastic effect. 
Introducing real and imaginary part, $\tilde{\chi}=\chi' - i\chi''$, we find 
\begin{align}
 \chi'(\freq)/\chi_0 & = \frac{1+q\tauKast(\tauB\freq)^2}{1+(\tauB\freq)^2} 
 - \frac{q \tauKast (\tau_{1}\freq)^2}{1+(\tau_1\freq)^2} 
 - \frac{2q(\tauB\freq)^2}{[1+(\tauB\freq)^2]^2} \label{chiprime}
\end{align}
and 
\begin{align}
 \chi''(\freq)/\chi_0 & = \frac{\tauB\freq(1 - q \tauKast)}{1+(\tauB\freq)^2} 
 + \frac{q \tauKast (\tau_{1}\freq)}{1+(\tau_1\freq)^2} 
 + \frac{q\tauB\freq(1-(\tauB\freq)^2)}{[1+(\tauB\freq)^2]^2} \label{chidprime}
\end{align}
From Eq.~(\ref{chidprime}) we find that the imaginary part $\chi''$ is no longer given by a single Lorentzian as in the Debye model. 
In qualitative agreement with experimental observations \cite{Roeben:2014co,Remmer:2017gf}, 
the location of the loss peak moves towards lower frequencies as the influence of viscoelasticity increases. 
At the same time, the height of the peak decreases and the width increases. 
All these features are seen in the experiments \cite{Roeben:2014co,Remmer:2017gf} and are described by Eq.~(\ref{chidprime}). 
The same conclusions have been reached in Ref.~\cite{Rusakov:2017gi} with the help of an effective field approximation and 
numerical solutions of the moment system (\ref{dta0})-(\ref{dta2}).
We compare the analytical formula in Eqs.~(\ref{chiprime}) and (\ref{chidprime}) to results of Brownian dynamics simulations 
shown in Fig.~\ref{chi.fig}. 
We find that results are relatively insensitive to the precise value of $\tauKast$ between $0.5$ and $2.0$, whereas 
corresponding variation in the value of $q$ leads to significant changes in the susceptibilities. 
The analytical formulae (\ref{chiprime}), (\ref{chidprime}) we find to be accurate for $q\lesssim0.5$.

\subsection{Magnetic Nanorheology}
In Ref.~\cite{Roeben:2014co}, Roeben {\em et al.}~suggest to transfer the Germant-DiMarzio-Bishop (GDB) model for the 
dielectric permittivity to the magnetic susceptibility, thereby relating the magnetic to the mechanical response, 
\begin{equation} \label{chi-GDB2}
\frac{\tilde{\chi}(\omega)-\chi_{\infty}}{\chi_{0}-\chi_{\infty}} = \frac{1}{1+i\omega\taueff(\omega)} = \frac{1}{1+K\tilde{G}(\omega)} .
\end{equation}
In the last equation, we introduced the complex shear modulus $\tilde{G}$ from 
$\tilde{G}(\omega) = i\omega\tilde{\eta}(\omega)$ and related the complex viscosity $\tilde{\eta}$ 
to the effective relaxation time via $\taueff(\omega)= \tilde{\xi}(\omega)/(2\kb T) = K \tilde{\eta}(\omega)$, 
where $K = 4\pi a^{3}/(\kb T)$. 

The idea of magnetic nanorheology proposed in \cite{Roeben:2014co} in this context is to use measurements of the magnetic susceptibilities 
$\tilde{\chi}=\chi'-i\chi''$ in order to infer information on mechanical properties of the environment where the nanoparticles perform their rotational relaxation. 
Reformulating Eq.~(\ref{chi-GDB2}), we find 
$K\tilde{G}(\omega) = \frac{1}{\chi^{\ast}(\omega)}- 1$, 
where $\chi^{\ast}=\frac{\tilde{\chi}(\omega)-\chi_{\infty}}{\chi_{0}-\chi_{\infty}}=\chi'_{N}-i\chi''_{N}$ and 
$\chi'_{N}=(\chi'-\chi_{\infty})/(\chi_{0}-\chi_{\infty})$ and 
$\chi''_{N}=\chi''/(\chi_{0}-\chi_{\infty})$. 
Defining the storage ($G'$) and loss modulus ($G''$), $\tilde{G}=G'+iG''$, we arrive at 
\begin{align}
G_{\rm GDB}'(\omega) & = \frac{1}{K} \left[ \frac{\chi'_{N}}{(\chi'_{N})^{2} + (\chi''_{N})^{2}} - 1\right] \label{Gprime}\\
G_{\rm GDB}''(\omega) & = \frac{1}{K} \frac{\chi''_{N}}{(\chi'_{N})^{2} + (\chi''_{N})^{2}} .\label{Gdprime}
\end{align}
Note that Eqs.~(\ref{Gprime}) and (\ref{Gdprime}) correct typos in the corresponding Eqs.~(8) and (9) of Ref.~\cite{Roeben:2014co}. 
In the Debye limit, $\chi'_{N}=\chi'/\chi_{0}=1/[1+(\tauB\omega)^{2}]$ and 
$\chi''_{N}=\chi''/\chi_{0}=\tauB\omega/[1+(\tauB\omega)^{2}]$, we find from  
Eqs.~(\ref{Gprime}) and (\ref{Gdprime}) the known result $G'=0$ and 
$G''=\tauB\omega/K=\etas\omega$ of a Newtonian fluid.

\begin{figure}
 \includegraphics[width=0.49\textwidth]{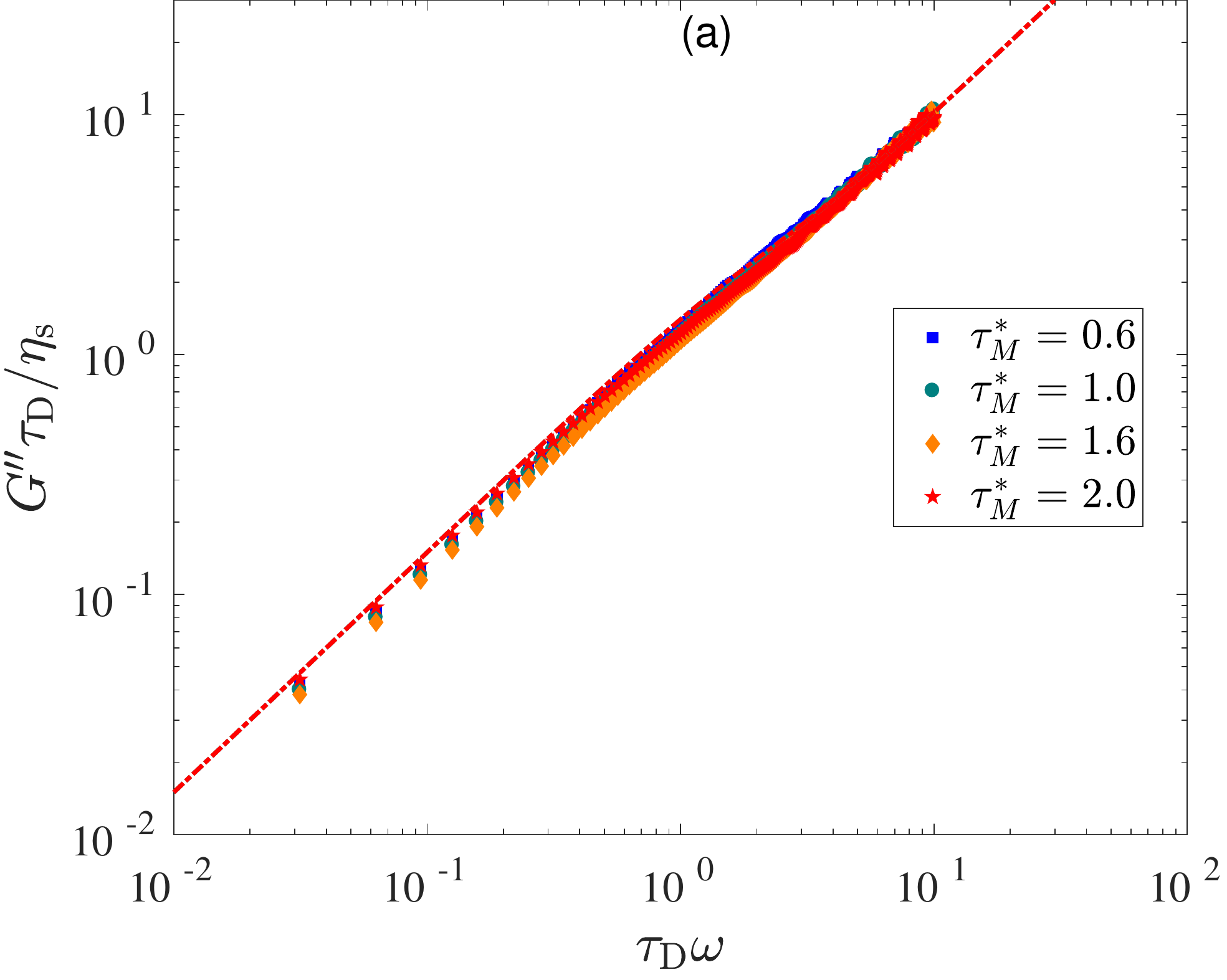}
 \includegraphics[width=0.49\textwidth]{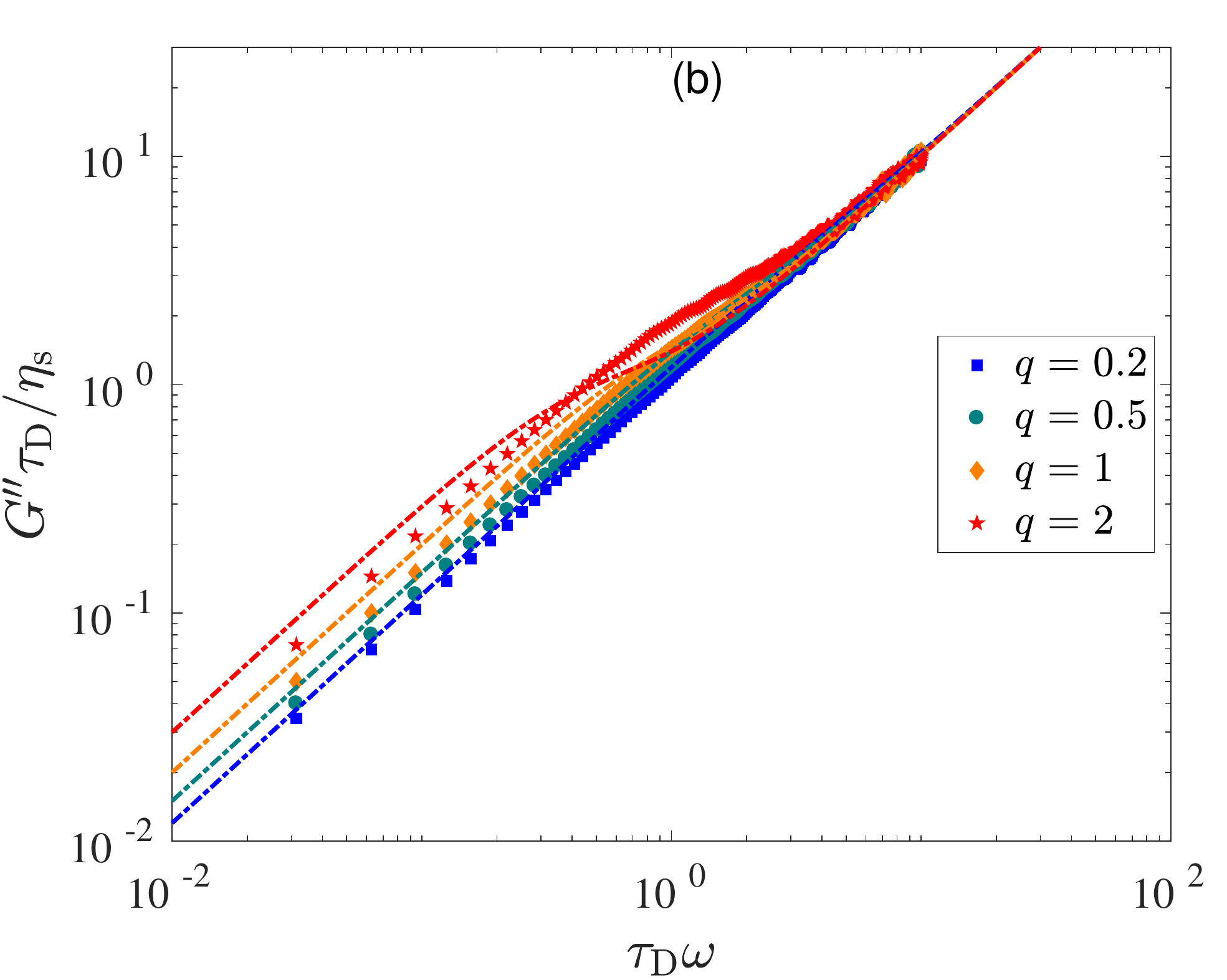}
 \caption{Left and right panel show the dimensionless loss modulus $G''(\omega)\tauB/\etas$ 
 as a function of $\tauB\omega$ for the same conditions as in Fig.~\ref{chi.fig}. 
 Symbols are results for the GDB assumption, Eq.~(\ref{GdprimeGDB}), using the numerical results for the 
 susceptibilites. 
 Lines correspond to the generalised Maxwell model, Eq.~(\ref{GdprimeGLE}).}
 \label{Gdprime.fig}
\end{figure}

For the model we study here, we worked out the magnetic susceptibilities in Eqs.~(\ref{chiprime}) and (\ref{chidprime}) 
to leading order in the ratio of friction coefficients $q$. 
Inserting these expressions into (\ref{Gprime}) and (\ref{Gdprime}), noting that $\chi_{\infty}=0$, we arrive at 
the GDB model for the storage and loss modulus, 
\begin{align}
G_{\rm GDB}'(\omega) & = \etas\omega \frac{q\tauK\omega}{(1+\tauK)^{2}+(\tauK\omega)^{2}}  + {\cal O}(q^{2}) \label{GprimeGDB}\\
G_{\rm GDB}''(\omega) & = \etas\omega \left[ 1 + q \frac{1+\tauKast}{(1+\tauKast)^{2}+(\tauK\omega)^{2}}\right] + {\cal O}(q^{2}) .\label{GdprimeGDB}
\end{align}
On the other hand, we can directly test Eqs.~(\ref{GprimeGDB}) and (\ref{GdprimeGDB}) by comparing them to the 
true mechanical modulus corresponding to the model under study. 
Since we assumed two independent frictional torques for the viscous and viscoelastic contribution we have 
$\tilde{\xi}_{\rm eff}(\omega) = \xi + \tilde{\zeta}(\omega)$. 
The retarded friction of the viscoelastic contribution was taken to be of the form $\zeta(t)=(\zeta_{0}/\tauK) e^{-t/\tauK}$ 
which leads to 
$\tilde{\zeta}(\omega)=\zeta_{0}/[1+i\omega\tauK]$ and therefore the effective relaxation time 
\begin{equation}
\tilde{\tau}_{\rm eff}(\omega) = \frac{\tilde{\xi}(\omega)}{2kT} = \tauB\left[ 1 + \frac{q}{1+i\tauK\omega} \right]
\end{equation}
Using the above definition of the complex modulus $\tilde{G}(\omega)=i\omega\tilde{\tau}_{\rm eff}(\omega)/K$, 
we find 
that the effective storage and loss modulus of the environment is given by 
\begin{align}
G'(\omega) & = \etas\omega  \frac{q\tauK\omega}{1+(\tauK\omega)^{2}}  \label{GprimeGLE}\\
G''(\omega) & =\etas\omega \left[ 1 + \frac{q}{1+(\tauK\omega)^{2}} \right] \label{GdprimeGLE}
\end{align}
These expressions for the generalized Maxwell (sometimes called Jeffrey's) model are similar but different from Eqs.~(\ref{GprimeGDB}), (\ref{GdprimeGDB}). 
We note that viscoelasticity effects vanish for $\tauK\to0$ and we recover from both expressions the results for a Newtonian solvent with 
and additional factor $1+q$ to account for the increased friction. 
For a general viscoelastic contribution, 
both expressions converge in the high-frequency limit to those of a Newtonian fluid as 
$\lim_{\omega\to\infty}G''(\omega)=G_{\infty}[1 + c/(\tauK\omega)^2]$, 
where $G_{\infty}=\etas\omega$ is the infinite-frequency shear modulus 
and $c=1+q$ and $c=q(1+\tauKast)$ for the present model and the GDB assumption, respectively. 
Viscoelasticity leads to a non-zero value of the high-frequency storage modulus 
$\lim_{\omega\to\infty}G'(\omega)=q/(K\tauKast)$. Higher order terms in $q$ appear in the GDB expression but not in Eq.~(\ref{GprimeGLE}). 
For low frequencies we find $\lim_{\omega\to 0}G''(\omega)=\etas\omega[1+bq]\to 0$ where $b=1$ and $b=1/(1+\tauKast)$ 
for the present model and the GDB assumption, respectively. 
The storage modulus vanishes quadratically for low frequencies, 
$\lim_{\omega\to 0}G'(\omega)=(q/K)\tauKast b^{2}(\tauB\omega)^{2}$. 
For low values of $q$ for which Eqs.~(\ref{GprimeGDB}), (\ref{GdprimeGDB}) apply, good agreement is found between the two expressions. 
This is especially true for the loss modulus, whereas the storage modulus shows a somewhat larger discrepancy (not shown). 
Figure \ref{Gdprime.fig} shows a comparison of $G''(\omega)$ as given by Eqs.~(\ref{GdprimeGLE}) 
and (\ref{GdprimeGDB}) for larger values of $q$. 
Due to the relative insensitivity of the susceptibilites on $\tauKast$ for intermediate values of $q$, 
also the value of $G''$ is only weakly affected and can be reliably estimated from the GDB model in this regime 
(see Fig.~\ref{Gdprime.fig}a). 
We find, however, some discrepancies for intermediate frequencies when the parameter $q$ increases. 
But overall, we find that the GDB model 
estimates the loss modulus for the present range of parameters quite well.  

\subsection{Magnetoviscosity}
The magnetoviscous effect, i.e.~the change of apparent viscosity due to an externally applied magnetic field, 
is not only of great theoretical interest but plays also an important role for various applications \cite{Ilg_lnp}. 
How is the magnetoviscous effect altered when the carrier medium is viscoelastic? 

To address this question, consider the situation where the system is exposed to a constant magnetic field $\bH$. 
In addition, a steady shear flow ${\bf V}=\dot{\gamma} y {\bf e}_{x}$ is applied, so that the vorticity 
is constant and 
$\bOmega = -(\dot{\gamma}/2) {\bf e}_{z}$. 
The nonequilibrium stationary state attained under the action of a constant magnetic field and $\bOmega$ is 
not known in general. 
For weak flow, $|\tauB\bOmega| = \tauB\dot{\gamma}/2 \ll 1$, we make the following ansatz 
\begin{equation} \label{f_ansatz}
f(\bu,\bz) = f_{0}(\bu,\bz)[ 1 + (\bu-\ave{\bu}_{0})\cdot\ba + \bz\cdot\bb + {\cal O}(\dot{\gamma}^{2})]
\end{equation}
where $f_{0}(\bu,\bz)$ is the equilibrium probability density (\ref{feq-uz}) in the presence of a magnetic field but 
in the absence of flow. 
The ansatz (\ref{f_ansatz}) satisfies the normalisation condition $\int\!\dd\bu\dd\bz\, f(\bu,\bz)=1$. 
The unknown vectors $\ba, \bb$ are independent of $\bu$ and $\bz$ and are assumed to be first 
order in the flow rate and therefore small. 
Otherwise positivity of $f$ is not guaranteed. 
As similar procedure has been used successfully for structure-forming ferrofluids \cite{IKHZ03}.

With the ansatz (\ref{f_ansatz}), we can calculate arbitrary moments in terms of $\ba, \bb$, e.g. 
\begin{align}
\ave{\bu} 
& = L_{1}\hbh + (L_{2}-L_{1}^{2})\hbh\hbh\cdot\ba + \frac{L_{1}}{h}\ba \label{ufromansatz}
\end{align}
where $\bh = h\hbh$ with $\bh^{2}=h^{2}$ and $\hbh$ the unit vector in the direction of the external field. 
Here we use $L_{j}(h)=\ave{P_{j}(\bu\cdot\hbh)}_{0}$, with 
$L_{1}(h)  = \coth(h) - 1/h$ the Langevin function introduced above and 
$L_{2}(h)  = 1 - 3L_1(h)/h$. 
We also find $\ave{\bz}=\zeq^2\bb$ and 
$\ave{\bo}=-L_1\zeq\hbh\times\bb$. 

From the Fokker-Planck equation, we can derive the following time evolution equations for the lowest order moments 
\begin{align}
\tauB \frac{\dd}{\dd t}\ave{\bu} & = - \ave{\bu} + \frac{1}{2}(\bh - \ave{\bu\bu}\cdot\bh) + \epsilon\ave{\bo} + \tauB\bOmega\times\ave{\bu}\label{duave}\\
\tauK \frac{\dd}{\dd t}\ave{\bz} & = -(1+q)\ave{\bz} - q\kb T\ave{\bu}\times\bh \label{dzave}
\end{align}
In the stationary state, the left hand side is zero. 
Expressing the moments  with the help of the ansatz (\ref{f_ansatz}) we arrive at an algebraic system of equations for 
the unknown $\ba$ and $\bb$. 
Solving this system of equations (see Appendix \ref{append:MVE}) and 
inserting the result into (\ref{ufromansatz}) we find the steady-state orientation due to an external field and weak flow as 
\begin{equation} \label{u-firstorder}
\ave{\bu} = L_{1}\hbh + \frac{L_{1}^{2}}{\frac{h}{2}\left(\frac{2+L_{2}}{3} -  \frac{qL_{1}^{2}}{1+p}\right)} \tauB \bOmega\times\hbh .
\end{equation}
For $q\to 0$, Eq.~(\ref{u-firstorder}) reduces to the known result of Shliomis model of ferrofluids \cite{Ilg_lnp}. 
The magnetization component perpendicular to the field direction leads to a viscous torque that manifests itself in 
the rotational viscosity \cite{rosensweigbook,Ilg_lnp}
\begin{equation}
\eta_{\rm rot} = \frac{M^{\perp}H}{2\dot{\gamma}} 
= \frac{M_{\rm sat}\kb T\tauB}{2m} \frac{\ave{u^{\perp}}h}{\tauB\dot{\gamma}} = (1/2)n\kb T\tauB  \frac{\ave{u^{\perp}}h}{\tauB\dot{\gamma}}
\end{equation}
Using $\tauB=\xi/(2kT)$ we find 
$n\kb T\tauB=n\xi/2=4n\pi\etas a^{3}=3\phi\etas$ with $\phi = n (4/3)\pi a^{3}$ the volume fraction. 
With these relations we arrive at the following expression for the rotational viscosity  
\begin{equation} \label{etarot}
\eta_{\rm rot} 
=  \frac{3}{2}\etas\phi\ \frac{3L_{1}^{2}}{2+L_{2}} \left[1 -  \frac{3qL_{1}^{2}}{(1+q)(2+L_{2})} \right]^{-1}.
\end{equation}
Therefore, we find that to first order in $q$ and in the flow rate the magnetoviscosity is  independent of $\tauK$ and depends only on $q$. 
The maximum viscosity contribution is 
\begin{equation} 
\eta_{\rm rot}^{\infty} = \lim_{h\to\infty}\eta_{\rm rot} =  \frac{3}{2}\etas\phi (1+q).
\end{equation}
Thus, the maximum viscosity increase due to a viscoelastic bath is lager by a factor $1+q$ then the pure viscous case 
simply by the increased rotational friction that the colloid experiences. 

For vanishing magnetic field we find $\eta_{\rm rot}(h=0)=0$, i.e.~no viscous contribution in the absence of a magnetic field. 
For weak fields $h\ll 1$, we find that the rotational viscosity increases as 
\begin{align}
\eta_{\rm rot} 
& = \frac{3}{2}\etas\phi\ \left(\frac{h^{2}}{6} - \frac{h^{4}}{36(1+q)}   + {\cal O}(h^{6}) \right) 
\end{align}

\begin{figure}[htb]
 \includegraphics[width=0.49\textwidth]{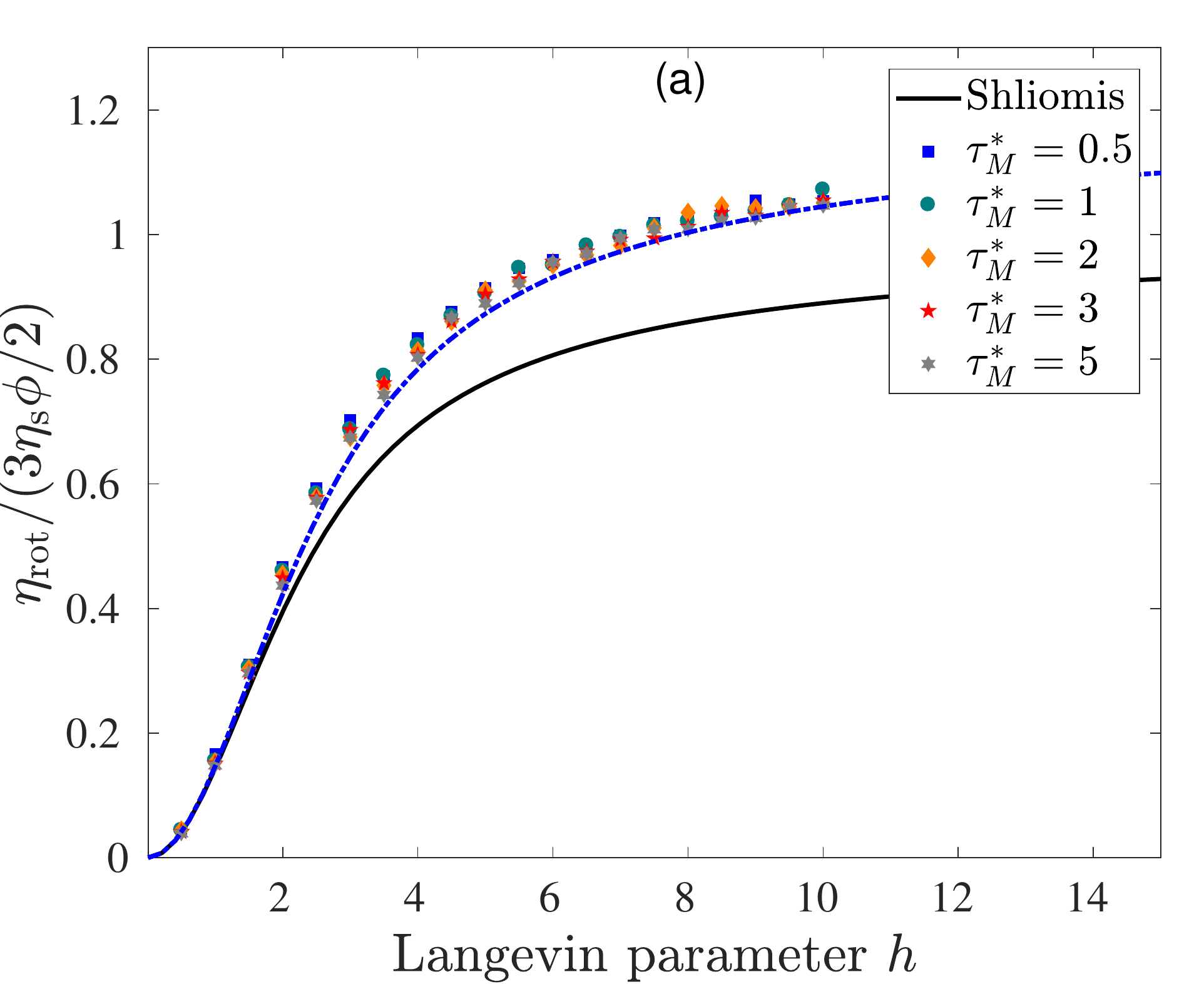}
 \includegraphics[width=0.49\textwidth]{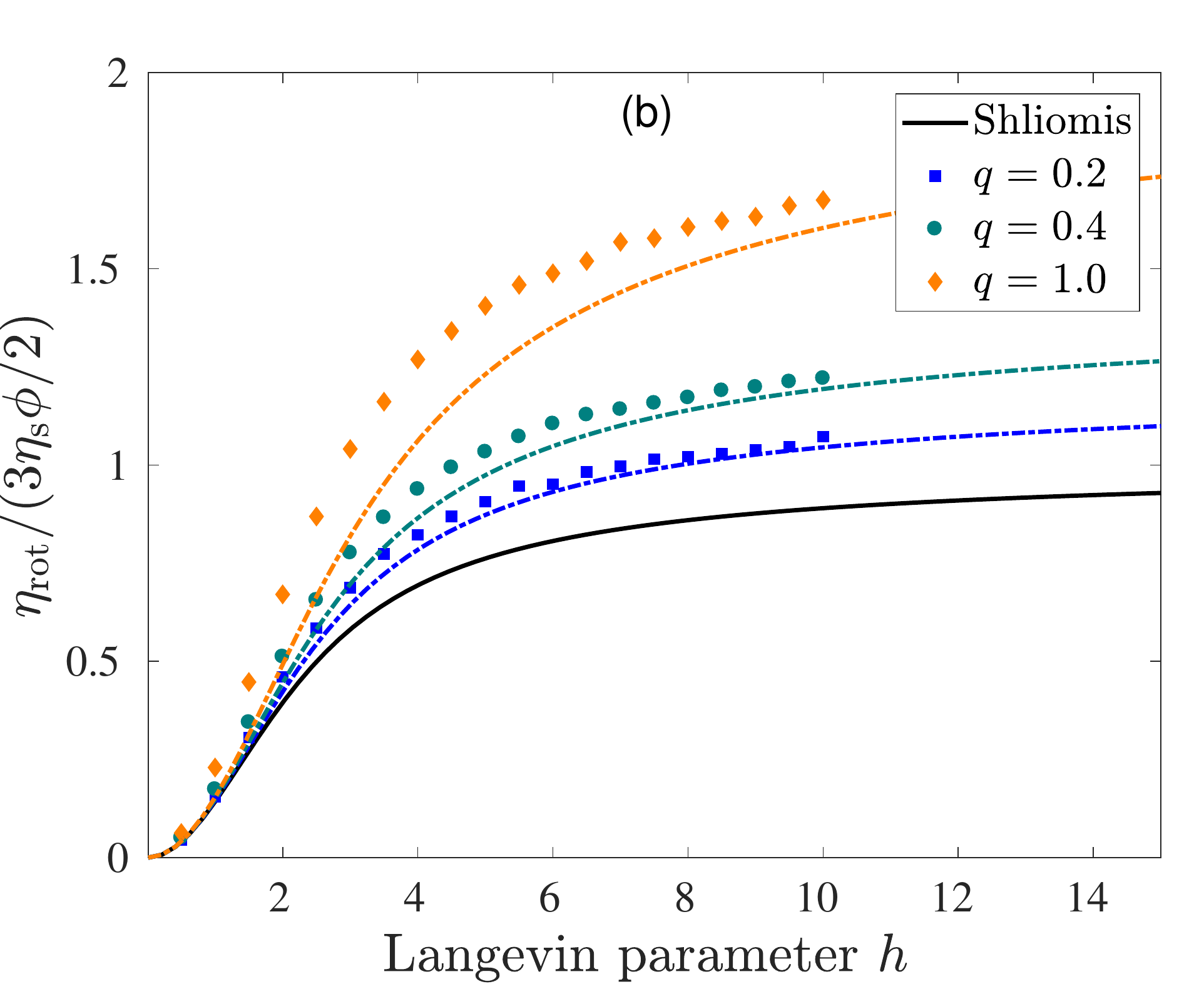}
 \caption{Scaled rotational viscosity $\eta_{\rm rot}/(3\etas\phi/2)$ as a function of the Langevin parameter $h$. 
 Symbols denote the results of Brownian dynamics simulations and dashed lines are 
 the analytical results (\ref{etarot}). The solid black line is the result for vanishing viscoelasticity. 
 Panel (a): Fixed value $q=0.2$ and $\tauKast$ varying from $0.5$ up to $5$. 
 Panel (b): From bottom to top $q=0.2, 0.4, 1.0$ with $\tauKast=1.0$.}
 \label{etarot.fig}
\end{figure}

\begin{figure}[htb]
 \includegraphics[width=0.49\textwidth]{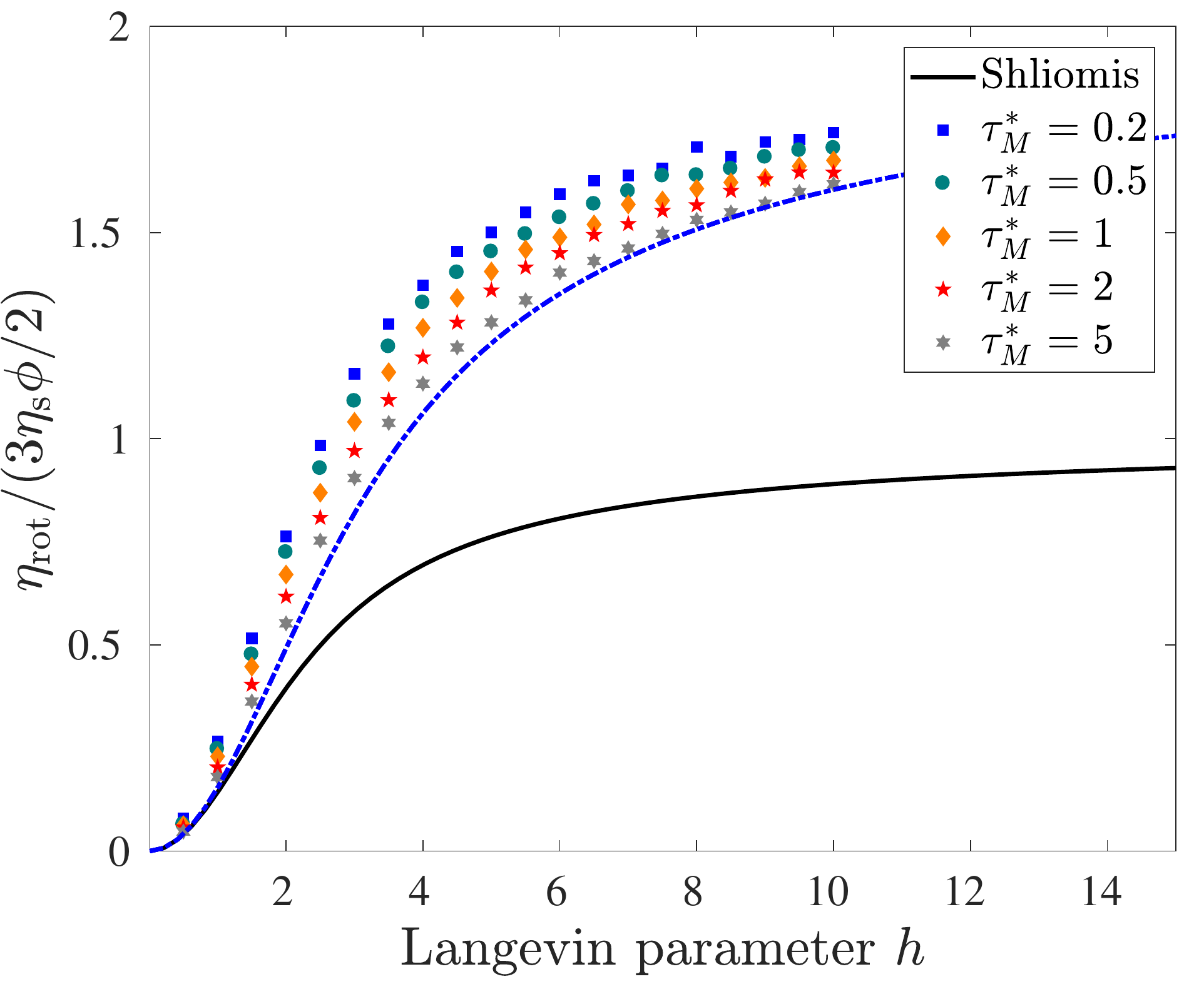}
 \caption{Same as right panel of Fig.~\ref{etarot.fig} but for $q=1$.}
 \label{etarot-p1.fig}
\end{figure}

Figure \ref{etarot.fig} shows the rotational viscosity as a function of the dimensionless applied magnetic field $h$. 
We find that the rotational viscosity increases proportionally to the additional viscoelastic friction. 
The value of $\tauKast$, on the contrary, has little effect on the steady state rotational viscosity in this parameter range. 
For comparison, we also performed Brownian dynamics simulations of Eqs.~(\ref{BDu}), (\ref{BDz}) with $\tauB\dot{\gamma}=0.1$. 
We verified that identical results are obtained for lower values of $\tauB\dot{\gamma}$. 
The numerical results from Brownian dynamics simulations are in good agreement with Eq.~(\ref{etarot}) for 
small values of $q$. 
For larger values of $q$, the analytical result (\ref{etarot}) significantly underestimates the true value 
of the rotational viscosity. 
Also the dependence of $\eta_{\rm rot}$ on the viscoelastic relaxation time $\tauK$ is not captured by Eq.~(\ref{etarot}). 
From Fig.~\ref{etarot-p1.fig}, we clearly see that increasing $\tauKast$ leads to a corresponding decrease of $\eta_{\rm rot}$. 
The same qualitative trend was found in Ref.~\cite{Raikher_MVEwithGLE} where only two-dimensional rotations in a purely viscoelastic bath were considered. 

\section{Conclusions} \label{concl.sec}
In the present contribution, we study a microscopic model of the non-Markovian dynamics of magnetic nanocolloids 
in a viscoelastic environment that can be described by the combination of a Newtonian and viscoelastic medium 
proposed in Ref.~\cite{Rusakov:2017gi}. 
When viewed as an extension of the basic model of ferrofluid dynamics, the additional viscoelasticity effects 
are described in this model by two dimensionless parameters: (i) the ratio $q=\zeta_0/\xi$ of friction coefficients  
of the nanocolloid in the viscoelastic and in the viscous component and (ii) the corresponding 
ratio $\tauKast=\tauK/\tauB$ of relaxation times. 
Due to the inherent non-linearity of three-dimensional rotational motion, we are only able to find analytical solutions 
of the model for weak viscoelasticity, i.e.~small values of $q$. 
We also test the analytical results against Brownian Dynamics simulations. 

We find that viscoelasticity leads to a slowing down of the magnetization relaxation compared to the purely viscous case, 
showing a non-exponential decay that is mainly controlled by the ratio $q$ of friction coefficients and 
rather insensitive to the precise value of scaled relaxation times $\tauKast$. 
Consequently, the magnetic susceptibility deviates from the Debye law, where the peak of $\chi''$ moves towards 
lower frequencies, while broadening and the amplitude decreasing as viscoelastic effects increase. 
These findings are in qualitative agreement with experimental results \cite{Roeben:2014co,Remmer:2017gf}. 
We also test a recent proposal put forward in Ref.~\cite{Roeben:2014co} to use measurements of the magnetic 
susceptibilities $\chi'$ and $\chi''$ to infer mechanical properties of the surrounding medium via the 
Germant-DiMarzio-Bishop relation. 
For the present model, we find that the GDB relation is  satisfied to a good approximation, at least 
for the parameter range investigated here. 
Finally, we work out the influence of viscoelasticity on the magnetoviscous effect. 
The increase of rotational friction due to the viscoelastic contribution leads to a corresponding increase in 
the maximum rotational viscosity. Besides this, increasing the relaxation time of the viscoelastic component 
relative to the viscous one leads to a decrease of the rotational viscosity. 
A similar reduction of the magnetoviscosity with increasing $\tauKast$ was found in \cite{Raikher_MVEwithGLE} for a simplified model. 

It will be interesting to compare the present model more quantitatively with experimental results on magnetic susceptibility, nanorheology and magnetoviscosity. Such comparisons would on the one hand allow for more reliable interpretation of the experimental results and on the other hand stimulate improvements over the current model formulation. 

\begin{acknowledgments}
Valuable discussions with Frank Ludwig, Annette Schmidt, Andreas Tsch\"ope are gratefully acknowledged. 
We also gratefully acknowledge support from the German Science Foundation (DFG) within the priority program SPP1681 
``Field controlled particle matrix interactions: synthesis multiscale modelling and application of magnetic hybrid materials'' 
under grant no.~IL 122/1-1. PI was also supported by a EU FP7--MC--CIG Grant No.~631233. 
\end{acknowledgments}

\begin{appendix}

\section{Short-time dynamics} \label{app:Cshort}
The time derivative of the correlation function $C(t)=\aveeq{\bu(t)\cdot\bu(0)}$ can be expressed as 
$\dot{C}(t)=\aveeq{[\LL^{\dagger}\bu(t)]\cdot\bu(0)}$. 
We find from the definition (\ref{L0def}) and (\ref{Lhdef}) of the adjoint operator that 
\begin{equation} \label{Ldagger-u}
\tauB \LL^{\dagger}\bu = -\bu + \epsilon\bo + \frac{1}{2}(\bh - \bu\bu\cdot\bh) + \bOmega\times\bu,
\end{equation}
where $o_{\alpha}=-\epsilon_{\alpha\beta\gamma}u_{\beta}z^{\ast}_{\gamma}$. 
Therefore, in the absence of external flow, $\tauB\dot{C}(0)=-C(0)+\epsilon\aveeq{\bo\cdot\bu}=-1$ since averages are taken with respect to the equilibrium state (\ref{feq-uz}) for which $\aveeq{\bo\cdot\bu}=0$. 

The next order term is found from $\ddot{C}(t)=\aveeq{[(\LL^{\dagger})^{2}\bu(t)]\cdot\bu(0)}$. 
Applying $\LL^{\dagger}$ to Eq.~(\ref{Ldagger-u}), we need 
\begin{align*}
\tauB\LL_{0}^{\dagger} \bo & = - \frac{\tauB}{\tau_{1}}\bo + \epsilon\bq + 4 \epsilon\bu\\
\tauB\LL_{h}^{\dagger} \bo & = \frac{1}{2}(\bzast\cdot\bu)\bu\times\bh - \frac{1}{2}\bzast\cdot(\bu\times\bh) - \epsilon(\bh-\bu\bu\cdot\bh)\\
\tauB\LL^{\dagger} \bu\bu\cdot\bh & = -2\bu\bu\cdot\bh + \epsilon(u(\bo\cdot\bh)+\bo(\bu\cdot\bh))\\
& + \frac{1}{2}(\bu h^{2} - 2\bu(\bu\cdot\bh)^{2}+\bh(\bu\cdot\bh)) 
\end{align*}
where
$q_{\alpha} = \epsilon_{\alpha\beta\gamma}\epsilon_{\mu\beta\lambda}u_{\lambda}z^{\ast}_{\gamma}z^{\ast}_{\mu}$. 
Inserting these expressions into $\ddot{C}$ and taking equilibrium averages for which 
$\aveeq{\bq\cdot\bu}=-2$, we arrive at Eq.~(\ref{uucorrel}).
 
\section{Moment equations for relaxational dynamics} \label{append:Mrelax}
We start with the moment system (\ref{dta0})-(\ref{dta2}) and insert the moment expansion (\ref{series-epsilon}). 
Matching equal orders of $\epsilon$ we find for ${\cal O}(\epsilon^{0})$: 
\begin{align}
\tauB\dot{a}_{0}^{(0)} & = - a_{0}^{(0)} \label{dta00}\\
\tauB\dot{a}_{1}^{(0)} & = - \lambda_{1}a_{1}^{(0)} \\
\tauB\dot{a}_{2}^{(0)} & = - \lambda_{2}a_{2}^{(0)} - \frac{4}{\tauK^{\ast}} a_{0}^{(0)}  \label{dota20}
\end{align}
From Eq (\ref{dta00}) we find 
\begin{equation}
a_{0}^{(0)}(t) = a_{0}^{(0)}(0) e^{-t/\tauB} = e^{-t/\tauB} 
\end{equation}
due to the initial condition $a_{0}^{(0)}(0)=\ave{u_{z}}(0)=1$. 
This is the familiar case where viscoelastic effects are absent, $q=0$. 
Furthermore, $a_{1}^{(0)}(t) = 0$ if we start with the equilibrium for a strong field and Gaussian in $\bz$, $a_{1}^{(0)}(0) =0$. 
Inserting $a_{0}^{(0)}(t) = e^{-t/\tauB}$ in Eq.~(\ref{dota20}), we find 
\begin{equation}
a_{2}^{(0)}(t) = -2e^{-t/\tauB} +(q_{z}(0)+2)e^{-t/\tau_{2}} = -2e^{-t/\tauB} 
\end{equation}
where we used 
\[
a_{2}^{(0)}(0) = \ave{q_{z}}(0)=\aveeq{z_{3}^{\ast}(\bzast\cdot\bu)}-\aveeq{\bu (\bzast)^{2}} =
\aveeq{(z_{3}^{\ast})^{2}} - \aveeq{(\bzast)^{2}} =-2
\]
from equilibrium Gaussian averages with Eq.~(\ref{feq-uz}). 

Now look at the first order terms: 
\begin{align}
\tauB\dot{a}_{0}^{(1)} & = - a_{0}^{(1)} + a_{1}^{(0)}\\
\tauB\dot{a}_{1}^{(1)} & = - \lambda_{1}a_{1}^{(1)} + a_{2}^{(0)} + 4a_{0}^{(0)}
\end{align}
Since $a_{1}^{(0)}=0$ we have also $a_{0}^{(1)}=0$ and therefore no correction to the magnetization dynamics 
to first order in $\epsilon$. 
The solution to $a_{1}^{(1)}$ reads 
\begin{equation}
a_{1}^{(1)}(t) = 2\tauK^{\ast}[ e^{-t/\tauB} -  e^{-t/\tau_{1}} ]
\end{equation}
where $a_{1}^{(1)}(0)=0$.  

Now, finally, we can compute the correction to the simple exponential magnetization decay from 
\begin{equation}
\tauB\dot{a}_{0}^{(2)} = - a_{0}^{(2)} + a_{1}^{(1)}
\end{equation}
and find the relaxation to second order in $\epsilon$ given by Eq.~(\ref{Mrelax}).

\section{Flow-induced deviation of orientation} \label{append:MVE}
With the ansatz (\ref{f_ansatz}), we can calculate arbitrary moments in terms of $\ba, \bb$, e.g. 
\begin{align}
\ave{\bu} & = \ave{\bu}_{0} + (\ave{\bu\bu}_{0}-\ave{\bu}_{0}\ave{\bu}_{0})\cdot\ba + \ave{\bu\bz}_{0}\cdot\bb 
\end{align}
where $\bh = h\hbh$ with $\hbh$ the unit vector in the direction of the external field. 
The first and second moment of the orientations are calculated from 
\begin{align*}
\ave{\bu\bu} & = \ave{\bu\bu}_{0} + (\ave{\bu\bu\bu}_{0}-\ave{\bu\bu}_{0}\ave{\bu}_{0})\cdot\ba + {\ave{\bu\bu\bz}_{0}\cdot\bb} \\
\ave{\bu\bu}_{0} & = L_{2}\hbh\hbh + \frac{L_{1}}{h}\bone \\
\ave{\bu\bu\bu}_{0} & = L_{3}\hbh\hbh\hbh + \frac{L_{2}}{h}(\hbh\bone)_{\rm sym} \\
\Rightarrow \ave{\bu\bu} & = L_{2}\hbh\hbh + \frac{L_{1}}{h}\bone 
+ (L_{3}-L_{2}L_{1})(\hbh\cdot\ba )\hbh\hbh
+ \frac{L_{2}-L_{1}^{2}}{h}(\hbh\cdot\ba)\bone + \frac{L_{2}}{h}(\ba\hbh + \hbh\ba)
\end{align*}
where $L_1, L_2$ have been introduced in the main text and $L_3(h)=L_1(h)-5L_2(h)/h$. 

For later use, we also provide 
\begin{align*}
\ave{\bu\bu}\cdot\hbh & = (L_{2}+L_{1}/h)\hbh + (L_{3}-L_{2}L_{1} + \frac{2L_{2}-L_{1}^{2}}{h} )\apara \hbh
+ \frac{L_{2}}{h}\ba
\end{align*}
where we defined $\apara=\ba\cdot\hbh$.

From Eq (\ref{dzave}), we find 
\begin{align}
0 & =  -(1+q)  \zeq^{2}\bb  - q\kb T L_{1}\ba\times\hbh \\
\Rightarrow \bb & = - \frac{q\kb TL_{1}}{(1+q)\zeq^{2}}\ba\times\hbh\\
\Rightarrow \ave{\bo} & = \frac{q\kb TL_{1}^{2}}{(1+q)\zeq}\hbh\times (\ba\times\hbh)
\end{align}

Inserting these results into (\ref{duave}) we find the condition for the stationary state to be 
\begin{align} \label{duavezero}
0  = & -L_{1}\hbh - (L_{2}-L_{1}^{2})\apara \hbh - \frac{L_{1}}{h}\ba \nonumber\\
&
+ \frac{h}{2}\left( [1 -  L_{2} - L_{1}/h] - (L_{3}-L_{2}L_{1} + \frac{2L_{2}-L_{1}^{2}}{h})\apara \right)\hbh -\frac{L_{2}}{2}\ba\nonumber\\
& 
+ \epsilon  \frac{q\kb TL_{1}^{2}}{(1+q)\zeq}\hbh\times (\ba\times\hbh) + \tauB L_{1}\bOmega\times\hbh
\end{align}
Note that $\hbh\times (\ba\times\hbh)=\ba-\apara\hbh$. 
Scalar multiplication of the above equation by $\hbh$ yields a linear equation for $\apara$: 
\begin{align}
0  
 = & \left[ - (L_{2}-L_{1}^{2} + \frac{L_{1}}{h}) 
- \frac{h}{2}\left( L_{3}-L_{2}L_{1} + \frac{3L_{2}-L_{1}^{2}}{h}\right) \right]\apara 
\end{align}
and therefore $\apara = 0$ 
since $1 -  L_{2} - L_{1}/h = 2L_{1}/h$ and therefore $ -L_{1} + \frac{h}{2}[1 -  L_{2} - L_{1}/h] = 0$. 

Thus, we know that $\ba = \apara\hbh + \baperp = \baperp$ with $\baperp\cdot\hbh=0$. 
Applying the orthogonal projector $\bone-\hbh\hbh$ to Eq (\ref{duavezero}) we arrive at 
\begin{align*}
0 = & - \frac{L_{1}}{h}\baperp  -\frac{L_{2}}{2}\baperp + \epsilon  \frac{q\kb TL_{1}^{2}}{(1+q)\zeq}\baperp + \tauB L_{1}\bOmega\times\hbh\\
= & \left(-\frac{2+L_{2}}{6} + \epsilon  \frac{q\kb TL_{1}^{2}}{(1+q)\zeq} \right)\baperp + \tauB L_{1}\bOmega\times\hbh\\
\Rightarrow 
\baperp = & \frac{L_{1}}{\frac{2+L_{2}}{6} - \epsilon  \frac{q\kb TL_{1}^{2}}{(1+q)\zeq}} \tauB \bOmega\times\hbh
\end{align*}
Inserting this result into (\ref{ufromansatz}) we find the mean orientation due to field and weak flow as given in 
Eq.~(\ref{u-firstorder}). 

\end{appendix}


%


\begin{thebibliography}{33}%
\makeatletter
\providecommand \@ifxundefined [1]{%
 \@ifx{#1\undefined}
}%
\providecommand \@ifnum [1]{%
 \ifnum #1\expandafter \@firstoftwo
 \else \expandafter \@secondoftwo
 \fi
}%
\providecommand \@ifx [1]{%
 \ifx #1\expandafter \@firstoftwo
 \else \expandafter \@secondoftwo
 \fi
}%
\providecommand \natexlab [1]{#1}%
\providecommand \enquote  [1]{``#1''}%
\providecommand \bibnamefont  [1]{#1}%
\providecommand \bibfnamefont [1]{#1}%
\providecommand \citenamefont [1]{#1}%
\providecommand \href@noop [0]{\@secondoftwo}%
\providecommand \href [0]{\begingroup \@sanitize@url \@href}%
\providecommand \@href[1]{\@@startlink{#1}\@@href}%
\providecommand \@@href[1]{\endgroup#1\@@endlink}%
\providecommand \@sanitize@url [0]{\catcode `\\12\catcode `\$12\catcode
  `\&12\catcode `\#12\catcode `\^12\catcode `\_12\catcode `\%12\relax}%
\providecommand \@@startlink[1]{}%
\providecommand \@@endlink[0]{}%
\providecommand \url  [0]{\begingroup\@sanitize@url \@url }%
\providecommand \@url [1]{\endgroup\@href {#1}{\urlprefix }}%
\providecommand \urlprefix  [0]{URL }%
\providecommand \Eprint [0]{\href }%
\providecommand \doibase [0]{http://dx.doi.org/}%
\providecommand \selectlanguage [0]{\@gobble}%
\providecommand \bibinfo  [0]{\@secondoftwo}%
\providecommand \bibfield  [0]{\@secondoftwo}%
\providecommand \translation [1]{[#1]}%
\providecommand \BibitemOpen [0]{}%
\providecommand \bibitemStop [0]{}%
\providecommand \bibitemNoStop [0]{.\EOS\space}%
\providecommand \EOS [0]{\spacefactor3000\relax}%
\providecommand \BibitemShut  [1]{\csname bibitem#1\endcsname}%
\let\auto@bib@innerbib\@empty
\bibitem [{\citenamefont {Rosensweig}(1985)}]{rosensweigbook}%
  \BibitemOpen
  \bibfield  {author} {\bibinfo {author} {\bibfnamefont {R.~E.}\ \bibnamefont
  {Rosensweig}},\ }\href@noop {} {\emph {\bibinfo {title}
  {Ferrohydrodynamics}}}\ (\bibinfo  {publisher} {Cambridge University Press},\
  \bibinfo {address} {Cambridge},\ \bibinfo {year} {1985})\BibitemShut
  {NoStop}%
\bibitem [{\citenamefont {Kr{\"o}ger}\ \emph {et~al.}(2003)\citenamefont
  {Kr{\"o}ger}, \citenamefont {Ilg},\ and\ \citenamefont
  {Hess}}]{KrogerIlgHess_JPHYS03}%
  \BibitemOpen
  \bibfield  {author} {\bibinfo {author} {\bibfnamefont {M.}~\bibnamefont
  {Kr{\"o}ger}}, \bibinfo {author} {\bibfnamefont {P.}~\bibnamefont {Ilg}}, \
  and\ \bibinfo {author} {\bibfnamefont {S.}~\bibnamefont {Hess}},\ }\href@noop
  {} {\bibfield  {journal} {\bibinfo  {journal} {J. Phys.: Condens. Matter}\
  }\textbf {\bibinfo {volume} {15}},\ \bibinfo {pages} {S1403} (\bibinfo {year}
  {2003})}\BibitemShut {NoStop}%
\bibitem [{\citenamefont {Torres-D\'iaz}\ and\ \citenamefont
  {Rinaldi}(2014)}]{Torres-Diaz2014}%
  \BibitemOpen
  \bibfield  {author} {\bibinfo {author} {\bibfnamefont {I.}~\bibnamefont
  {Torres-D\'iaz}}\ and\ \bibinfo {author} {\bibfnamefont {C.}~\bibnamefont
  {Rinaldi}},\ }\href@noop {} {\bibfield  {journal} {\bibinfo  {journal} {Soft
  Matter}\ }\textbf {\bibinfo {volume} {10}},\ \bibinfo {pages} {8584}
  (\bibinfo {year} {2014})}\BibitemShut {NoStop}%
\bibitem [{\citenamefont {Holm}\ and\ \citenamefont
  {Weis}(2005)}]{Holm_review2005}%
  \BibitemOpen
  \bibfield  {author} {\bibinfo {author} {\bibfnamefont {C.}~\bibnamefont
  {Holm}}\ and\ \bibinfo {author} {\bibfnamefont {J.~J.}\ \bibnamefont
  {Weis}},\ }\href@noop {} {\bibfield  {journal} {\bibinfo  {journal} {Curr.
  Opin. Colloid Interface Sci.}\ }\textbf {\bibinfo {volume} {10}},\ \bibinfo
  {pages} {133} (\bibinfo {year} {2005})}\BibitemShut {NoStop}%
\bibitem [{\citenamefont {Ilg}\ and\ \citenamefont {Odenbach}(2008)}]{Ilg_lnp}%
  \BibitemOpen
  \bibfield  {author} {\bibinfo {author} {\bibfnamefont {P.}~\bibnamefont
  {Ilg}}\ and\ \bibinfo {author} {\bibfnamefont {S.}~\bibnamefont {Odenbach}},\
  }in\ \href@noop {} {\emph {\bibinfo {booktitle} {Colloidal Magnetic Fluids:
  Basics, Development and Applications of Ferrofluids}}},\ \bibinfo {series}
  {Lecture Notes in Phys.}, Vol.\ \bibinfo {volume} {763},\ \bibinfo {editor}
  {edited by\ \bibinfo {editor} {\bibfnamefont {S.}~\bibnamefont {Odenbach}}}\
  (\bibinfo  {publisher} {Springer},\ \bibinfo {address} {Berlin},\ \bibinfo
  {year} {2008})\BibitemShut {NoStop}%
\bibitem [{\citenamefont {Zr\'inyi}(2000)}]{Zrinyi_review}%
  \BibitemOpen
  \bibfield  {author} {\bibinfo {author} {\bibfnamefont {M.}~\bibnamefont
  {Zr\'inyi}},\ }\href@noop {} {\bibfield  {journal} {\bibinfo  {journal}
  {Colloid Polym. Sci}\ }\textbf {\bibinfo {volume} {278}},\ \bibinfo {pages}
  {98} (\bibinfo {year} {2000})}\BibitemShut {NoStop}%
\bibitem [{\citenamefont {Messing}\ \emph {et~al.}(2011)\citenamefont
  {Messing}, \citenamefont {Frickel}, \citenamefont {Belkoura}, \citenamefont
  {Strey}, \citenamefont {Rahn}, \citenamefont {Odenbach},\ and\ \citenamefont
  {Schmidt}}]{Annette_magnetlinkedgel}%
  \BibitemOpen
  \bibfield  {author} {\bibinfo {author} {\bibfnamefont {R.}~\bibnamefont
  {Messing}}, \bibinfo {author} {\bibfnamefont {N.}~\bibnamefont {Frickel}},
  \bibinfo {author} {\bibfnamefont {L.}~\bibnamefont {Belkoura}}, \bibinfo
  {author} {\bibfnamefont {R.}~\bibnamefont {Strey}}, \bibinfo {author}
  {\bibfnamefont {H.}~\bibnamefont {Rahn}}, \bibinfo {author} {\bibfnamefont
  {S.}~\bibnamefont {Odenbach}}, \ and\ \bibinfo {author} {\bibfnamefont
  {A.~M.}\ \bibnamefont {Schmidt}},\ }\href@noop {} {\bibfield  {journal}
  {\bibinfo  {journal} {Macromolecules}\ }\textbf {\bibinfo {volume} {44}},\
  \bibinfo {pages} {2990} (\bibinfo {year} {2011})}\BibitemShut {NoStop}%
\bibitem [{\citenamefont {Bender}\ \emph {et~al.}(2011)\citenamefont {Bender},
  \citenamefont {G\"unther}, \citenamefont {Tsch\"ope},\ and\ \citenamefont
  {Birringer}}]{Tschoepe_ferrogel2011}%
  \BibitemOpen
  \bibfield  {author} {\bibinfo {author} {\bibfnamefont {P.}~\bibnamefont
  {Bender}}, \bibinfo {author} {\bibfnamefont {A.}~\bibnamefont {G\"unther}},
  \bibinfo {author} {\bibfnamefont {A.}~\bibnamefont {Tsch\"ope}}, \ and\
  \bibinfo {author} {\bibfnamefont {R.}~\bibnamefont {Birringer}},\ }\href@noop
  {} {\bibfield  {journal} {\bibinfo  {journal} {J. Magn. Magn. Mater.}\
  }\textbf {\bibinfo {volume} {323}},\ \bibinfo {pages} {2055} (\bibinfo {year}
  {2011})}\BibitemShut {NoStop}%
\bibitem [{\citenamefont {Reddy}\ \emph {et~al.}(2011)\citenamefont {Reddy},
  \citenamefont {Mohan}, \citenamefont {Varaprasad}, \citenamefont {Ravindra},
  \citenamefont {Joy},\ and\ \citenamefont {Raju}}]{Reddy_ferrogel}%
  \BibitemOpen
  \bibfield  {author} {\bibinfo {author} {\bibfnamefont {N.~N.}\ \bibnamefont
  {Reddy}}, \bibinfo {author} {\bibfnamefont {Y.~M.}\ \bibnamefont {Mohan}},
  \bibinfo {author} {\bibfnamefont {K.}~\bibnamefont {Varaprasad}}, \bibinfo
  {author} {\bibfnamefont {S.}~\bibnamefont {Ravindra}}, \bibinfo {author}
  {\bibfnamefont {P.~A.}\ \bibnamefont {Joy}}, \ and\ \bibinfo {author}
  {\bibfnamefont {K.~M.}\ \bibnamefont {Raju}},\ }\href@noop {} {\bibfield
  {journal} {\bibinfo  {journal} {J. Appl. Polym. Sci.}\ }\textbf {\bibinfo
  {volume} {122}},\ \bibinfo {pages} {1364} (\bibinfo {year}
  {2011})}\BibitemShut {NoStop}%
\bibitem [{\citenamefont {Ilg}(2013)}]{pi_hydrogel}%
  \BibitemOpen
  \bibfield  {author} {\bibinfo {author} {\bibfnamefont {P.}~\bibnamefont
  {Ilg}},\ }\href@noop {} {\bibfield  {journal} {\bibinfo  {journal} {Soft
  Matter}\ }\textbf {\bibinfo {volume} {9}},\ \bibinfo {pages} {3465} (\bibinfo
  {year} {2013})}\BibitemShut {NoStop}%
\bibitem [{\citenamefont {Hergt}\ \emph {et~al.}(2006)\citenamefont {Hergt},
  \citenamefont {Dutz}, \citenamefont {M\"uller},\ and\ \citenamefont
  {Zeisberger}}]{Hergt2006}%
  \BibitemOpen
  \bibfield  {author} {\bibinfo {author} {\bibfnamefont {R.}~\bibnamefont
  {Hergt}}, \bibinfo {author} {\bibfnamefont {S.}~\bibnamefont {Dutz}},
  \bibinfo {author} {\bibfnamefont {R.}~\bibnamefont {M\"uller}}, \ and\
  \bibinfo {author} {\bibfnamefont {M.}~\bibnamefont {Zeisberger}},\
  }\href@noop {} {\bibfield  {journal} {\bibinfo  {journal} {J. Phys.: Condens.
  Matter}\ }\textbf {\bibinfo {volume} {18}},\ \bibinfo {pages} {S2919}
  (\bibinfo {year} {2006})}\BibitemShut {NoStop}%
\bibitem [{\citenamefont {Alexiou}\ \emph {et~al.}(2007)\citenamefont
  {Alexiou}, \citenamefont {Jurgons}, \citenamefont {Seliger}, \citenamefont
  {Brunke}, \citenamefont {Iro},\ and\ \citenamefont {Odenbach}}]{Alexiou2007}%
  \BibitemOpen
  \bibfield  {author} {\bibinfo {author} {\bibfnamefont {C.}~\bibnamefont
  {Alexiou}}, \bibinfo {author} {\bibfnamefont {R.}~\bibnamefont {Jurgons}},
  \bibinfo {author} {\bibfnamefont {C.}~\bibnamefont {Seliger}}, \bibinfo
  {author} {\bibfnamefont {O.}~\bibnamefont {Brunke}}, \bibinfo {author}
  {\bibfnamefont {H.}~\bibnamefont {Iro}}, \ and\ \bibinfo {author}
  {\bibfnamefont {S.}~\bibnamefont {Odenbach}},\ }\href@noop {} {\bibfield
  {journal} {\bibinfo  {journal} {Anticancer Research}\ }\textbf {\bibinfo
  {volume} {27}},\ \bibinfo {pages} {2019} (\bibinfo {year}
  {2007})}\BibitemShut {NoStop}%
\bibitem [{\citenamefont {Wiekhorst}\ \emph {et~al.}(2012)\citenamefont
  {Wiekhorst}, \citenamefont {Steinhoff}, \citenamefont {Eberbeck},\ and\
  \citenamefont {Trahms}}]{Wiekhorst:2012fz}%
  \BibitemOpen
  \bibfield  {author} {\bibinfo {author} {\bibfnamefont {F.}~\bibnamefont
  {Wiekhorst}}, \bibinfo {author} {\bibfnamefont {U.}~\bibnamefont
  {Steinhoff}}, \bibinfo {author} {\bibfnamefont {D.}~\bibnamefont {Eberbeck}},
  \ and\ \bibinfo {author} {\bibfnamefont {L.}~\bibnamefont {Trahms}},\
  }\href@noop {} {\bibfield  {journal} {\bibinfo  {journal} {Pharmaceutical
  research}\ }\textbf {\bibinfo {volume} {29}},\ \bibinfo {pages} {1189}
  (\bibinfo {year} {2012})}\BibitemShut {NoStop}%
\bibitem [{\citenamefont {Haun}\ \emph {et~al.}(2010)\citenamefont {Haun},
  \citenamefont {Yoon}, \citenamefont {Lee},\ and\ \citenamefont
  {Weissleder}}]{Haun:2010kg}%
  \BibitemOpen
  \bibfield  {author} {\bibinfo {author} {\bibfnamefont {J.~B.}\ \bibnamefont
  {Haun}}, \bibinfo {author} {\bibfnamefont {T.~J.}\ \bibnamefont {Yoon}},
  \bibinfo {author} {\bibfnamefont {H.}~\bibnamefont {Lee}}, \ and\ \bibinfo
  {author} {\bibfnamefont {R.}~\bibnamefont {Weissleder}},\ }\href@noop {}
  {\bibfield  {journal} {\bibinfo  {journal} {Wiley Interdisciplinary Reviews:
  Nanomedicine and Nanobiotechnology}\ }\textbf {\bibinfo {volume} {2}},\
  \bibinfo {pages} {291} (\bibinfo {year} {2010})}\BibitemShut {NoStop}%
\bibitem [{\citenamefont {Wilhelm}\ \emph {et~al.}(2003)\citenamefont
  {Wilhelm}, \citenamefont {Gazeau},\ and\ \citenamefont
  {Bacri}}]{Wilhelm2003}%
  \BibitemOpen
  \bibfield  {author} {\bibinfo {author} {\bibfnamefont {C.}~\bibnamefont
  {Wilhelm}}, \bibinfo {author} {\bibfnamefont {F.}~\bibnamefont {Gazeau}}, \
  and\ \bibinfo {author} {\bibfnamefont {J.-C.}\ \bibnamefont {Bacri}},\
  }\href@noop {} {\bibfield  {journal} {\bibinfo  {journal} {Phys. Rev. E}\
  }\textbf {\bibinfo {volume} {67}},\ \bibinfo {pages} {061908} (\bibinfo
  {year} {2003})}\BibitemShut {NoStop}%
\bibitem [{\citenamefont {Barrera}\ \emph {et~al.}(2010)\citenamefont
  {Barrera}, \citenamefont {Flori{\'a}n-Algarin}, \citenamefont {Acevedo},\
  and\ \citenamefont {Rinaldi}}]{Barrera:2010kf}%
  \BibitemOpen
  \bibfield  {author} {\bibinfo {author} {\bibfnamefont {C.}~\bibnamefont
  {Barrera}}, \bibinfo {author} {\bibfnamefont {V.}~\bibnamefont
  {Flori{\'a}n-Algarin}}, \bibinfo {author} {\bibfnamefont {A.}~\bibnamefont
  {Acevedo}}, \ and\ \bibinfo {author} {\bibfnamefont {C.}~\bibnamefont
  {Rinaldi}},\ }\href@noop {} {\bibfield  {journal} {\bibinfo  {journal} {Soft
  Matter}\ }\textbf {\bibinfo {volume} {6}},\ \bibinfo {pages} {3662} (\bibinfo
  {year} {2010})}\BibitemShut {NoStop}%
\bibitem [{\citenamefont {Remmer}\ \emph {et~al.}(2017)\citenamefont {Remmer},
  \citenamefont {Roeben}, \citenamefont {Schmidt}, \citenamefont {Schilling},\
  and\ \citenamefont {Ludwig}}]{Remmer:2017gf}%
  \BibitemOpen
  \bibfield  {author} {\bibinfo {author} {\bibfnamefont {H.}~\bibnamefont
  {Remmer}}, \bibinfo {author} {\bibfnamefont {E.}~\bibnamefont {Roeben}},
  \bibinfo {author} {\bibfnamefont {A.~M.}\ \bibnamefont {Schmidt}}, \bibinfo
  {author} {\bibfnamefont {M.}~\bibnamefont {Schilling}}, \ and\ \bibinfo
  {author} {\bibfnamefont {F.}~\bibnamefont {Ludwig}},\ }\href@noop {}
  {\bibfield  {journal} {\bibinfo  {journal} {Journal of Magnetism and Magnetic
  Materials}\ }\textbf {\bibinfo {volume} {427}},\ \bibinfo {pages} {331}
  (\bibinfo {year} {2017})}\BibitemShut {NoStop}%
\bibitem [{\citenamefont {Tsch{\"o}pe}\ \emph {et~al.}(2014)\citenamefont
  {Tsch{\"o}pe}, \citenamefont {Birster}, \citenamefont {Trapp}, \citenamefont
  {Bender},\ and\ \citenamefont {Birringer}}]{Tschope:2014kj}%
  \BibitemOpen
  \bibfield  {author} {\bibinfo {author} {\bibfnamefont {A.}~\bibnamefont
  {Tsch{\"o}pe}}, \bibinfo {author} {\bibfnamefont {K.}~\bibnamefont
  {Birster}}, \bibinfo {author} {\bibfnamefont {B.}~\bibnamefont {Trapp}},
  \bibinfo {author} {\bibfnamefont {P.}~\bibnamefont {Bender}}, \ and\ \bibinfo
  {author} {\bibfnamefont {R.}~\bibnamefont {Birringer}},\ }\href@noop {}
  {\bibfield  {journal} {\bibinfo  {journal} {Journal Of Applied Physics}\
  }\textbf {\bibinfo {volume} {116}},\ \bibinfo {pages} {184305} (\bibinfo
  {year} {2014})}\BibitemShut {NoStop}%
\bibitem [{\citenamefont {Roeben}\ \emph {et~al.}(2014)\citenamefont {Roeben},
  \citenamefont {Roeder}, \citenamefont {Teusch}, \citenamefont {Effertz},
  \citenamefont {Deiters},\ and\ \citenamefont {Schmidt}}]{Roeben:2014co}%
  \BibitemOpen
  \bibfield  {author} {\bibinfo {author} {\bibfnamefont {E.}~\bibnamefont
  {Roeben}}, \bibinfo {author} {\bibfnamefont {L.}~\bibnamefont {Roeder}},
  \bibinfo {author} {\bibfnamefont {S.}~\bibnamefont {Teusch}}, \bibinfo
  {author} {\bibfnamefont {M.}~\bibnamefont {Effertz}}, \bibinfo {author}
  {\bibfnamefont {U.~K.}\ \bibnamefont {Deiters}}, \ and\ \bibinfo {author}
  {\bibfnamefont {A.~M.}\ \bibnamefont {Schmidt}},\ }\href@noop {} {\bibfield
  {journal} {\bibinfo  {journal} {Colloid and Polymer Science}\ }\textbf
  {\bibinfo {volume} {292}},\ \bibinfo {pages} {2013} (\bibinfo {year}
  {2014})}\BibitemShut {NoStop}%
\bibitem [{\citenamefont {Jarkova}\ \emph {et~al.}(2003)\citenamefont
  {Jarkova}, \citenamefont {Pleiner}, \citenamefont {M{\"u}ller},\ and\
  \citenamefont {Brand}}]{Pleiner_ferrogel}%
  \BibitemOpen
  \bibfield  {author} {\bibinfo {author} {\bibfnamefont {E.}~\bibnamefont
  {Jarkova}}, \bibinfo {author} {\bibfnamefont {H.}~\bibnamefont {Pleiner}},
  \bibinfo {author} {\bibfnamefont {H.-W.}\ \bibnamefont {M{\"u}ller}}, \ and\
  \bibinfo {author} {\bibfnamefont {H.}~\bibnamefont {Brand}},\ }\href@noop {}
  {\bibfield  {journal} {\bibinfo  {journal} {Phys. Rev. E}\ }\textbf {\bibinfo
  {volume} {68}},\ \bibinfo {pages} {041706} (\bibinfo {year}
  {2003})}\BibitemShut {NoStop}%
\bibitem [{\citenamefont {Attaran}\ \emph {et~al.}(2017)\citenamefont
  {Attaran}, \citenamefont {Brummund},\ and\ \citenamefont
  {Wallmersperger}}]{Attaran2017}%
  \BibitemOpen
  \bibfield  {author} {\bibinfo {author} {\bibfnamefont {A.}~\bibnamefont
  {Attaran}}, \bibinfo {author} {\bibfnamefont {J.}~\bibnamefont {Brummund}}, \
  and\ \bibinfo {author} {\bibfnamefont {T.}~\bibnamefont {Wallmersperger}},\
  }\href@noop {} {\bibfield  {journal} {\bibinfo  {journal} {Journal of
  Intelligent Material Systems and Structures}\ }\textbf {\bibinfo {volume}
  {28}},\ \bibinfo {pages} {1358} (\bibinfo {year} {2017})}\BibitemShut
  {NoStop}%
\bibitem [{\citenamefont {Pessot}\ \emph {et~al.}(2015)\citenamefont {Pessot},
  \citenamefont {Weeber}, \citenamefont {Holm}, \citenamefont {L{\"o}wen},\
  and\ \citenamefont {Menzel}}]{Pessot2015}%
  \BibitemOpen
  \bibfield  {author} {\bibinfo {author} {\bibfnamefont {G.}~\bibnamefont
  {Pessot}}, \bibinfo {author} {\bibfnamefont {R.}~\bibnamefont {Weeber}},
  \bibinfo {author} {\bibfnamefont {C.}~\bibnamefont {Holm}}, \bibinfo {author}
  {\bibfnamefont {H.}~\bibnamefont {L{\"o}wen}}, \ and\ \bibinfo {author}
  {\bibfnamefont {A.~M.}\ \bibnamefont {Menzel}},\ }\href@noop {} {\bibfield
  {journal} {\bibinfo  {journal} {J. Phys.: Condens. Matter}\ }\textbf
  {\bibinfo {volume} {27}},\ \bibinfo {pages} {325105} (\bibinfo {year}
  {2015})}\BibitemShut {NoStop}%
\bibitem [{\citenamefont {Volkov}\ and\ \citenamefont
  {Leonov}(2001)}]{Volkov:2001jr}%
  \BibitemOpen
  \bibfield  {author} {\bibinfo {author} {\bibfnamefont {V.~S.}\ \bibnamefont
  {Volkov}}\ and\ \bibinfo {author} {\bibfnamefont {A.~I.}\ \bibnamefont
  {Leonov}},\ }\href@noop {} {\bibfield  {journal} {\bibinfo  {journal}
  {Physical Review E}\ }\textbf {\bibinfo {volume} {64}},\ \bibinfo {pages}
  {051113} (\bibinfo {year} {2001})}\BibitemShut {NoStop}%
\bibitem [{\citenamefont {Raikher}\ and\ \citenamefont
  {Rusakov}(2005)}]{Raikher2005}%
  \BibitemOpen
  \bibfield  {author} {\bibinfo {author} {\bibfnamefont {Y.~L.}\ \bibnamefont
  {Raikher}}\ and\ \bibinfo {author} {\bibfnamefont {V.~V.}\ \bibnamefont
  {Rusakov}},\ }\href@noop {} {\bibfield  {journal} {\bibinfo  {journal}
  {Colloid Journal}\ }\textbf {\bibinfo {volume} {67}},\ \bibinfo {pages} {610}
  (\bibinfo {year} {2005})}\BibitemShut {NoStop}%
\bibitem [{\citenamefont {Raikher}\ and\ \citenamefont
  {Rusakov}(2008)}]{Raikher_MVEwithGLE}%
  \BibitemOpen
  \bibfield  {author} {\bibinfo {author} {\bibfnamefont {Y.~L.}\ \bibnamefont
  {Raikher}}\ and\ \bibinfo {author} {\bibfnamefont {V.~V.}\ \bibnamefont
  {Rusakov}},\ }\href@noop {} {\bibfield  {journal} {\bibinfo  {journal}
  {Colloid Journal}\ }\textbf {\bibinfo {volume} {70}},\ \bibinfo {pages} {77}
  (\bibinfo {year} {2008})}\BibitemShut {NoStop}%
\bibitem [{\citenamefont {Raikher}\ \emph {et~al.}(2013)\citenamefont
  {Raikher}, \citenamefont {Rusakov},\ and\ \citenamefont
  {Perzynski}}]{Raikher:2013ju}%
  \BibitemOpen
  \bibfield  {author} {\bibinfo {author} {\bibfnamefont {Y.~L.}\ \bibnamefont
  {Raikher}}, \bibinfo {author} {\bibfnamefont {V.~V.}\ \bibnamefont
  {Rusakov}}, \ and\ \bibinfo {author} {\bibfnamefont {R.}~\bibnamefont
  {Perzynski}},\ }\href@noop {} {\bibfield  {journal} {\bibinfo  {journal}
  {Soft Matter}\ }\textbf {\bibinfo {volume} {9}},\ \bibinfo {pages} {10857}
  (\bibinfo {year} {2013})}\BibitemShut {NoStop}%
\bibitem [{\citenamefont {Rusakov}\ and\ \citenamefont
  {Raikher}(2017)}]{Rusakov:2017gi}%
  \BibitemOpen
  \bibfield  {author} {\bibinfo {author} {\bibfnamefont {V.~V.}\ \bibnamefont
  {Rusakov}}\ and\ \bibinfo {author} {\bibfnamefont {Y.~L.}\ \bibnamefont
  {Raikher}},\ }\href@noop {} {\bibfield  {journal} {\bibinfo  {journal}
  {Journal of Chemical Physics}\ }\textbf {\bibinfo {volume} {147}} (\bibinfo
  {year} {2017})}\BibitemShut {NoStop}%
\bibitem [{\citenamefont {Mazenko}(2006)}]{Mazenko}%
  \BibitemOpen
  \bibfield  {author} {\bibinfo {author} {\bibfnamefont {G.}~\bibnamefont
  {Mazenko}},\ }\href@noop {} {\emph {\bibinfo {title} {Nonequilibrium
  Statistical Mechanics}}}\ (\bibinfo  {publisher} {Wiley-VCH},\ \bibinfo
  {address} {Weinheim},\ \bibinfo {year} {2006})\BibitemShut {NoStop}%
\bibitem [{\citenamefont {Grimm}\ \emph {et~al.}(2011)\citenamefont {Grimm},
  \citenamefont {Jeney},\ and\ \citenamefont {Franosch}}]{Grimm:2011ei}%
  \BibitemOpen
  \bibfield  {author} {\bibinfo {author} {\bibfnamefont {M.}~\bibnamefont
  {Grimm}}, \bibinfo {author} {\bibfnamefont {S.}~\bibnamefont {Jeney}}, \ and\
  \bibinfo {author} {\bibfnamefont {T.}~\bibnamefont {Franosch}},\ }\href@noop
  {} {\bibfield  {journal} {\bibinfo  {journal} {Soft Matter}\ }\textbf
  {\bibinfo {volume} {7}},\ \bibinfo {pages} {2076} (\bibinfo {year}
  {2011})}\BibitemShut {NoStop}%
\bibitem [{\citenamefont {Dhont}(1996)}]{Dhont_book}%
  \BibitemOpen
  \bibfield  {author} {\bibinfo {author} {\bibfnamefont {J.~K.~G.}\
  \bibnamefont {Dhont}},\ }\href@noop {} {\emph {\bibinfo {title} {An
  introduction to dynamics of colloids}}},\ Studies in interface science\
  (\bibinfo  {publisher} {Elsevier},\ \bibinfo {address} {Amsterdam},\ \bibinfo
  {year} {1996})\BibitemShut {NoStop}%
\bibitem [{\citenamefont {Fannin}\ \emph {et~al.}(1995)\citenamefont {Fannin},
  \citenamefont {Charles},\ and\ \citenamefont {Relihan}}]{Fannin95}%
  \BibitemOpen
  \bibfield  {author} {\bibinfo {author} {\bibfnamefont {P.~C.}\ \bibnamefont
  {Fannin}}, \bibinfo {author} {\bibfnamefont {S.~W.}\ \bibnamefont {Charles}},
  \ and\ \bibinfo {author} {\bibfnamefont {T.}~\bibnamefont {Relihan}},\
  }\href@noop {} {\bibfield  {journal} {\bibinfo  {journal} {J. Phys. D: Appl.
  Phys.}\ }\textbf {\bibinfo {volume} {28}},\ \bibinfo {pages} {1765} (\bibinfo
  {year} {1995})}\BibitemShut {NoStop}%
\bibitem [{\citenamefont {{\"O}ttinger}(1996)}]{hcobook}%
  \BibitemOpen
  \bibfield  {author} {\bibinfo {author} {\bibfnamefont {H.~C.}\ \bibnamefont
  {{\"O}ttinger}},\ }\href@noop {} {\emph {\bibinfo {title} {Stochastic
  Processes in Polymeric Fluids}}}\ (\bibinfo  {publisher} {Springer},\
  \bibinfo {address} {Berlin},\ \bibinfo {year} {1996})\BibitemShut {NoStop}%
\bibitem [{\citenamefont {Ilg}\ \emph {et~al.}(2003)\citenamefont {Ilg},
  \citenamefont {Kr{\"o}ger}, \citenamefont {Hess},\ and\ \citenamefont
  {Zubarev}}]{IKHZ03}%
  \BibitemOpen
  \bibfield  {author} {\bibinfo {author} {\bibfnamefont {P.}~\bibnamefont
  {Ilg}}, \bibinfo {author} {\bibfnamefont {M.}~\bibnamefont {Kr{\"o}ger}},
  \bibinfo {author} {\bibfnamefont {S.}~\bibnamefont {Hess}}, \ and\ \bibinfo
  {author} {\bibfnamefont {A.~Y.}\ \bibnamefont {Zubarev}},\ }\href@noop {}
  {\bibfield  {journal} {\bibinfo  {journal} {Phys. Rev. E}\ }\textbf {\bibinfo
  {volume} {67}},\ \bibinfo {pages} {061401} (\bibinfo {year}
  {2003})}\BibitemShut {NoStop}%
\end{thebibliography}
\end{document}